\documentclass[twocolumn]{aastex631}
\usepackage{graphicx}
\usepackage{amsmath}
\usepackage{hyperref}

\submitjournal{\apj}
\received{XXX}
\revised{YYY}
\accepted{ZZZ}

\usepackage{color}

\defcitealias{peeples19}{Paper~I}
\defcitealias{corlies20}{Paper~II}
\defcitealias{zheng20}{Paper~III}
\defcitealias{simons20}{Paper~IV}
\defcitealias{lochhaas21}{Paper~V}
\defcitealias{lochhaas23}{Paper~VI}

\shorttitle{(Dis)Assembling FOGGIE Stellar Halos}
\shortauthors{A. C. Wright et al.}

\graphicspath{{./}{Figures/}}

\begin{document}
\title{Figuring Out Gas \& Galaxies In Enzo (FOGGIE) VII: The (Dis)Assembly of Stellar Halos}
\correspondingauthor{Anna C.\ Wright}
\email{acwright@jhu.edu}

\author[0000-0002-1685-5818]{Anna C.\ Wright}
\affiliation{Center for Astrophysical Sciences, William H.\ Miller III Department of Physics \& Astronomy, Johns Hopkins University, 3400 N.\ Charles Street, Baltimore, MD 21218}

\author[0000-0002-7982-412X]{Jason Tumlinson}
\affiliation{Space Telescope Science Institute, 3700 San Martin Dr., Baltimore, MD 21218}
\affiliation{Center for Astrophysical Sciences, William H.\ Miller III Department of Physics \& Astronomy, Johns Hopkins University, 3400 N.\ Charles Street, Baltimore, MD 21218}

\author[0000-0003-1455-8788]{Molly S.\ Peeples}
\affiliation{Space Telescope Science Institute, 3700 San Martin Dr., Baltimore, MD 21218}
\affiliation{Center for Astrophysical Sciences, William H.\ Miller III Department of Physics \& Astronomy, Johns Hopkins University, 3400 N.\ Charles Street, Baltimore, MD 21218}

\author[0000-0002-2786-0348]{Brian W.\ O'Shea}
\affiliation{Department of Computational Mathematics, Science, and Engineering, 
Department of Physics and Astronomy, 
National Superconducting Cyclotron Laboratory,  
Michigan State University}

\author[0000-0003-1785-8022]{Cassandra Lochhaas}
\affiliation{Space Telescope Science Institute, 3700 San Martin Dr., Baltimore, MD 21218}
\affiliation{Center for Astrophysics, Harvard \& Smithsonian, 60 Garden St., Cambridge, MA 02138}
\affiliation{NASA Hubble Fellow}

\author[0000-0002-0646-1540]{Lauren Corlies}
\affiliation{Adler Planetarium, 1300 S DuSable Lake Shore Dr., Chicago, IL 60605}

\author[0000-0002-6804-630X]{Britton D.\ Smith}
\affiliation{Institute for Astronomy, University of Edinburgh, Royal Observatory, EH9 3HJ, UK}

\author[0009-0009-9700-1811]{Nguyen Binh}
\affiliation{Department of Astronomy and Steward Observatory, University of Arizona, Tucson, AZ 85721, USA}
\affiliation{Space Telescope Science Institute, 3700 San Martin Dr., Baltimore, MD 21218}
\affiliation{Astronomy Department, University of Washington, Seattle, WA 98195, USA}

\author[0000-0001-7472-3824]{Ramona Augustin}
\affiliation{Space Telescope Science Institute, 3700 San Martin Dr., Baltimore, MD 21218}

\author[0000-0002-6386-7299]{Raymond C.\ Simons}
\affiliation{Department of Physics, University of Connecticut, Storrs, CT 06269 USA}
\affiliation{Space Telescope Science Institute, 3700 San Martin Dr., Baltimore, MD 21218}

\begin{abstract}
%No more than 250 words 
Over the next decade, the astronomical community will be commissioning multiple wide-field observatories well-suited for studying stellar halos in both integrated light and resolved stars. In preparation for this, we use five high-resolution cosmological simulations of Milky Way-like galaxies from the FOGGIE suite to explore the properties and components of stellar halos. These simulations are run with high time (5\,Myr) and stellar mass (1000\,M$_\odot$) resolution to better model the properties and origins of low density regions like stellar halos. We find that the FOGGIE stellar halos have masses, metallicity gradients, and surface brightness profiles that are consistent with observations. In agreement with other simulations, the FOGGIE stellar halos receive 30--40\% of their mass from in situ stars. However, this population is more centrally concentrated in the FOGGIE simulations and therefore does not contribute excess light to the halo outskirts. The remaining stars are accreted from $\sim 10$--50 other galaxies, with the majority of the accreted mass originating in 2–4 galaxies. While the inner halo ($r<50$\,kpc) of each FOGGIE galaxy has a large number of contributors, the halo outskirts of three of the five galaxies are primarily made up of stars from only a few contributors. We predict that upcoming wide-field observatories, like the \textit{Nancy Grace Roman Space Telescope}, will probe stellar halos around Milky Way-like galaxies out to $\sim 100$\,kpc in integrated light and will be able to distinguish the debris of dwarf galaxies with extended star formation histories from the underlying halo with resolved color-magnitude diagrams.
\end{abstract}

% \keywords{}

\section{Introduction}
\label{intro}
Observations of the stellar populations of Milky Way-like galaxies are almost always limited to the galaxy's disk and brightest satellites \citep[e.g.,][]{Tollerud2011,Geha2017,Carlsten2022}. However, this is only a small part of a much larger whole. Observations that probe lower luminosities, fainter surface brightnesses, and/or wider fields have revealed that the majority of Milky Way-like galaxies are surrounded by not only large populations of faint dwarf galaxies and globular clusters, but also extended and diffuse structures of stars known as stellar halos \citep[e.g.,][]{Mouhcine2005,Merritt2016,Harmsen2017}. Although these components contain only a small fraction of the galaxy's total stellar mass, they preserve a detailed record of the early universe and the assembly history of the system as a whole.

In the $\Lambda$CDM model of cosmic structure formation, galaxies like the Milky Way are built from the mergers of many much smaller objects \citep[e.g.,][]{White1978,Searle1978}. While the bulk of the baryonic material from this assembly process ultimately becomes part of the disk, stars that are stripped from infalling satellites \citep[e.g.,][]{Johnston1995,Ibata2001} or perturbed during mergers \citep[e.g.,][]{Zolotov2009,Purcell2010} frequently adopt non-disk orbits that may take them far from the center of the galaxy. Low densities at large galactocentric distances result in long dynamical times, allowing structures in the stellar halo to persist for sometimes billions of years \citep[e.g.,][]{Johnston1996}. The global properties of the stellar halo and the distribution of substructure within it therefore contain information about the mass accretion history of the central galaxy and the many galaxies that have contributed to the system over time \citep[e.g.,][]{Johnston1998,Helmi1999b,Johnston2008,Amorisco2017}.

Because of their proximity, the stellar halos of the Milky Way and M31 have thus far been our primary sources of data. Both halos are predominantly old and metal-poor \citep[e.g.,][]{Unavane1996,Chiba2000,Kalirai2006,Carollo2007}, but have been found to contain a substantial amount of substructure \citep{Ibata1994,Majewski1996,Chiba2000,Ivezic2000,Yanny2000,Ibata2001b,Newberg2002,Ferguson2002,Ibata2007,Juric2008,Bell2008,Gilbert2012,Naidu2020}, some of which is chemically distinct from the bulk of the halo \citep[e.g.,][]{Cohen2018,Deason2023}. Although stars that likely originated in the central disks have been found in the stellar halos of both the Milky Way and M31 \citep[e.g.,][]{Carollo2007,Bonaca2017}, the general consensus is that most of their stars formed in satellites that were eventually tidally disrupted \citep[e.g.,][]{Searle1978,Majewski1996,Bullock2001,Purcell2007,Bell2008}, and differences between the halos can therefore generally be attributed to differences in their accretion histories. The nearly 100 stellar streams that have been discovered in the Milky Way halo are evidence that our galaxy has accreted and disrupted many dwarf galaxies and globular clusters throughout its history \citep[e.g.,][]{Belokurov2006,Shipp2018,Ibata2019}. However, the unbroken density profile of M31's halo, combined with its higher mass, higher metallicity, and steeper metallicity gradient suggest that M31 has had a considerably more active and extended accretion history than the Milky Way \citep[e.g.,][]{Deason2013,Gilbert2014}.

Our ability to measure the detailed kinematics and chemical compositions of these nearby halos has also allowed us to derive substantial information about the population of galaxies that produced them. For instance, differences between the abundances found in Milky Way halo stars and present-day satellites were initially thought to rule out destroyed dwarf galaxies as the main contributors to the halo \citep[][]{Unavane1996,Venn2004}. However, this disparity is now thought to indicate that the surviving classical satellites of the Milky Way are a biased subset of the many companions our galaxy has had throughout its history. While present-day dwarfs were typically accreted fairly recently, the primary building blocks of the halo were early infalling dwarfs that assembled relatively close to the Milky Way. These early contributors to the stellar halo formed their stars rapidly and therefore with little enrichment from Type Ia supernovae, making them $\alpha$-enhanced relative to surviving dwarfs \citep[e.g.,][]{Robertson2005,Corlies2013,Fattahi2020,Naidu2022}. 

Many authors \citep[e.g.,][]{Helmi1999a,Johnston2001,Lee2015,Li2022} have also performed detailed decompositions of stellar halos to associate stars with individual accretion events and thereby derive the likely infall times, star formation histories, and masses of individual contributors. \citet{Naidu2020} recently used kinematic and chemical data from the H3 Survey \citep{Conroy2019a} and the \textit{Gaia} mission \citep{GaiaCollaboration2018} to associate $\geq$95\% of a sample of giant stars in the inner ($r<50$\,kpc) Milky Way stellar halo with specific structures, the vast majority of which are believed to be the remnants of disrupted dwarfs. In M31, data from the Pan-Andromeda Archaeological Survey \citep[PAndAS;][]{McConnachie2009}, Project AMIGA \citep[Absorption Maps in the Gas of Andromeda;][]{Cohen2018}, the Spectroscopic and Photometric Landscape of Andromeda’s Stellar Halo \citep[SPLASH;][]{Gilbert2009}, and other surveys have been used to derive the likely properties of the progenitor of the Giant Stellar Stream \citep[GSS; e.g.,][]{Ibata2001b,Conn2016,D'Souza2018b,Gilbert2019}. 

Beyond the Local Group, the large sizes and low surface brightnesses of stellar halos make them particularly challenging observational targets. Wide-field ground-based observations have been used to map the distributions of giant stars in a number of more distant stellar halos \citep[e.g.,][]{Mouhcine2010,Bailin2011,Barker2012,Crnojevic2013,Okamoto2015,Smercina2023,Gozman2023}, often in combination with pencil-beam surveys of overlapping fields imaged by \textit{HST}. Other studies have relied on integrated light to detect brighter substructure like streams and shells \citep[e.g.,][]{Martinez-Delgado2015,Merritt2016}. These observations have shown that stellar halos around other spiral galaxies have much in common with the Milky Way and M31's halos: most consist primarily of old stars \citep[e.g,][]{Mouhcine2005b,Monachesi2013} and many feature relatively bright stellar streams \citep[e.g.,][]{Malin1997,Shang1998,Martinez-Delgado2010,Hood2018}. However, there is also considerable diversity in the mass, luminosity, extent, and radial trends of these halos \citep{Merritt2016,Harmsen2017,Gilhuly2022} that is thought to result from their varying accretion histories \citep[e.g.,][]{Smercina2020,Smercina2022}. For instance, the metallicities of stellar halos appear to reflect the mass of their most significant contributor. The strong mass-metallicity trend observed in dwarf galaxies means that more massive stellar halos, which typically have more massive contributors, are more metal-rich than less massive halos \citep[e.g.,][]{Mouhcine2005,Harmsen2017,D'Souza2018}. Observations of diverse stellar halos therefore provide a broader context for the differences that we observe between the Milky Way and M31's halos. However, the sample size of observed stellar halos remains relatively small, and observations of only the brightest features or small patches of resolved stars are inherently limited.

Over the next decade, the astronomical community will be commissioning multiple wide-field observatories that have the potential to observe stellar halos in much greater detail and much larger numbers than ever before \citep{Johnston2001}. The Vera C. Rubin Observatory \citep{LSSTScienceCollaboration2009}, \textit{Euclid} \citep{Laureijs2011}, and the \textit{Nancy Grace Roman Space Telescope} \citep{Spergel2015} will image large patches of the sky at significant depth and in a variety of wavelengths. \textit{Roman}, in particular, combines a wide field-of-view with the sensitivity and resolution necessary to resolve individual stars in stellar halos out to $D\geq10$\,Mpc \citep{Lancaster2022}. This means that we are entering an era where we will have access to the kind of detailed data previously only achievable in the Local Group over a much larger volume. We will have the ability to consider the resolved stellar populations of stellar halos as a standard part of the larger picture, rather than as an exception. 

On the theory side, simulations run in support of these observations must be able to model low surface brightness regions with high fidelity. In this paper, we analyze a suite of high-resolution zoom-in simulations of Milky Way-like galaxies and their stellar halos. The FOGGIE simulations are run with unusually high time and stellar mass resolution, which, as we will demonstrate, are assets in analyzing the properties and origins of stellar halos. We discuss the FOGGIE simulations and our stellar halo selection method in \S\,\ref{sims} and \S\,\ref{kinid}, respectively, then explore how the properties of the FOGGIE halos compare to observed stellar halos in \S\,\ref{props}. In \S\,\ref{dis}, we look at the characteristics and distributions of the contributors to the FOGGIE stellar halos and in \S\,\ref{futsur} we examine the implications of our findings for future wide-field surveys.

\section{The FOGGIE Simulation Suite}
\label{sims}
The galaxies we analyze in this paper are from the FOGGIE (Figuring Out Gas \& Galaxies in Enzo) simulation suite. The FOGGIE simulations were first introduced in \citet{Peeples2019} and \citet{Corlies2020}, with the runs we analyze here discussed in \citet{Simons2020} and \citet{Lochhaas2021}. We summarize the properties of these simulations in \S\,\ref{sec:halos}, the unique refinement scheme we use and its importance in  \S\,\ref{sec:refine}, the physical prescriptions for the UV background, star formation, and feedback in \S\,\ref{sec:radiation}--\ref{sec:feedback}, and our methods for subhalo finding in \S\,\ref{sec:halofinding}. We also include a brief discussion of the caveats of the simulations in \S\,\ref{caveats}.

\subsection{Halo Selection}\label{sec:halos}
The FOGGIE production simulations consist of six high-resolution cosmological simulations of Milky Way-like galaxies run with the adaptive mesh refinement (AMR) code \textsc{Enzo} \citep{Bryan2014,Brummel-Smith2019}\footnote{\href{http://enzo-project.org}{http://enzo-project.org}}. In \textsc{Enzo}, the gravitational potential is computed via the Particle-Mesh method on the root grid and a multigrid Poisson solver on adaptively-refined grids, using a total density field calculated from all particles (stars and dark matter) and the gas density field. Gas is evolved by solving Euler's equations of hydrodynamics on the grid using the piecewise parabolic method (PPM). Each simulation is evolved to $z=0$ using a flat $\Lambda$CDM cosmology with $1-\Omega_\Lambda = \Omega_\mathrm{m}=0.285$, $\Omega_\mathrm{b} = 0.0461$, and $h=0.695$. The five FOGGIE simulations that have reached $z=0$ are included in our analysis. 

The central galaxies of the FOGGIE simulations were drawn from a dark-matter-only run in a cubic domain 100 Mpc/$h$ on a side (in comoving coordinates). We selected these halos based on two criteria: 1) a $z=0$ virial mass similar to that estimated for the Milky Way \citep[$\sim$10$^{12}$\,M$_\odot$;][]{Bland-Hawthorn2016} and 2) no major (mass ratio $>$ 10:1) mergers after $z\approx2$, the time at which the Milky Way is thought to have experienced its last major merger \citep{Helmi2018}. Each galaxy is then re-simulated from $z=99$ at much higher resolution and with full hydrodynamics using the ``zoom-in'' technique. All dark matter particles located within $3R_\mathrm{vir}$ of the central galaxy at $z=0$ are re-simulated with a mass of $M_\mathrm{dm,part} = 1.39\times10^6 \, $M$_\odot$. The masses of the other dark matter particles increase with distance from the region of interest, up to a maximum of 5.69$\times$10$^9$ M$_\odot$ far away from the zoom region. 

\subsection{Natural, Forced, and Cooling Refinement}\label{sec:refine}

Throughout the majority of the simulation volume, the refinement of the underlying grid is determined by the local mass density. When the baryonic or dark matter mass of a cell exceeds a threshold mass corresponding to 8 times the mean mass of the highest resolution cells in the simulation initial conditions, the cell is divided in half along each dimension such that 
\begin{equation}
\ell_{\mathrm{cell}}(N_{\mathrm{ref}}) = 2^{-N_{\mathrm{ref}}}\times\frac{\ell_{\mathrm{box}}}{N_{\mathrm{root}}},
\end{equation}
where $\ell_{\mathrm{cell}}$ is the new length of the cell, $N_{\mathrm{ref}}$ is the level to which it is being refined, $\ell_{\mathrm{box}}$ is the length of the simulation box, and $N_{\mathrm{root}}$ is the number of root grid cells on a side. This ensures that, when this refinement criterion is dominant, grid cells maintain roughly similar masses across refinement levels. The FOGGIE simulations contain $256^3$ root grid cells and are permitted to refine up to 11 levels, so the minimum cell size in each simulation is 274 comoving pc (cpc). 

The FOGGIE simulations also employ additional refinement schemes to improve resolution in and around their central galaxies. Those low mass dark matter particles that lie within the zoom region are designated as ``must refine particles'' \citep{Simpson2013}. Using a cloud-in-cell algorithm, \textsc{Enzo} flags the cells nearest a given must refine particle and forces the cells to refine to a minimum of $N_{\mathrm{ref}}=4$ (35\,ckpc). However, a much more stringent resolution requirement is placed on the cells nearest to the central galaxy. Beginning at $z=6$, the simulations employ a ``forced refinement'' scheme, which consists of a (288\,ckpc)$^3$ box that tracks the center of mass of the central galaxy throughout the domain and enforces a minimum of 9 levels of refinement (1.10\,ckpc) within its boundaries. These cells may refine up to two more levels if the cooling length of the gas they contain is less than the length of the cell. 

The combination of the forced and cooling refinement schemes greatly improves the spatial and mass resolution in the warm and hot gas that fills most of the volume of the circumgalactic medium (CGM). In a traditional density-based refinement scheme, the interstellar medium (ISM) of the galaxy would reach $N_\mathrm{ref}=11$ (274\,cpc), while the CGM would only reach $N_\mathrm{ref}=6$--8 (2.20--8.78\,ckpc; \citealp{Corlies2020}). By enforcing a higher level of refinement in the CGM, the FOGGIE simulations are able to reduce artificial mixing and resolve the detailed kinematics of the CGM \citep{Peeples2019,Corlies2020,Lochhaas2021}. The cooling refinement scheme further improves the resolution of thermally unstable gas, resulting in the cooling length being resolved in $>99$\% of the CGM by volume and $>90$\% of the CGM by mass at $z=2$ \citep{Simons2020}.

The effects of these refinement schemes on the CGM have already been explored in FOGGIE Papers I--VI \citep{Peeples2019,Corlies2020,Zheng2020,Simons2020,Lochhaas2021,Lochhaas2023}. However, these techniques are also relevant to studies of the stellar halo because of their effects on satellites. \citet{Simons2020} showed that the wide range of density and velocity structures in the FOGGIE CGM causes satellites to experience ram pressure that varies over five orders of magnitude as they orbit the central galaxy. Ram pressure stripping is also highly stochastic, with 90\% of the total surface momentum from ram pressure being imparted in less than 20\% of a satellite's orbital time at $z\geq2$. As a result, satellites in the FOGGIE simulations experience less ram pressure stripping on average compared to satellites in a spherically averaged hydrostatic CGM. The resolution of the CGM may therefore affect the formation of stars in dwarfs that contribute to the stellar halo.

\subsection{Cooling and Background Radiation}\label{sec:radiation}

In order to approximate the effects of reionization, the FOGGIE simulations include a redshift-dependent UV background following \citet{Haardt2012} with HI self-shielding following \citet{Emerick2019}. The simulations compute primordial cooling by solving a non-equilibrium chemical reaction network for H, H$^+$, He, He$^+$, He$^{++}$, and e$^-$ using the \textsc{Grackle} chemical and cooling library \citep{Smith2017}. All metal species are grouped together, so the FOGGIE simulations also include metallicity-dependent cooling assuming ionization equilibrium and solar abundances.

\subsection{Star Formation}\label{sec:sf}

Star formation within the simulations is based on the properties of local gas and broadly follows \citet{Cen1992} and \citet{Cen2006}. Star particles form from gas that fulfills the following criteria (\textsc{Enzo StarParticleCreation} method 0):
\begin{enumerate}
\item the gas cell does not have a higher AMR level inside of it and the gas density exceeds $10^4\times$ the mean density of all matter within the simulation ($n\geq0.016$\, cm$^{-3}$ at $z=0$), 
\item the divergence of the gas cell's velocity is negative, 
\item the cooling time of the gas cell is less than its dynamical time or the temperature of the gas is $<$11,000 K, 
\item the gas cell is Jeans unstable, and 
\item the gas cell contains sufficient mass that the star particle that would form from it would exceed a minimum mass threshold. 
\end{enumerate}
In the smallest of the FOGGIE simulations, Tempest, the minimum star particle mass is held at 1000 M$_\odot$ over the full run to $z=0$. However, in the other simulations, this criterion varies with time. The initial value is 1000\,M$_\odot$, but the threshold increases linearly with time to 10$^4$\,M$_\odot$ between $z=2$ and $z=1$. This means that regions composed primarily of old stellar populations, like the stellar halo, end up with higher mass resolution than e.g., the younger spiral arms. 

If all of the requirements are met, a star particle will form at the center of the gas cell with
\begin{equation}
M_{\mathrm{\star,part}} = c^\star M_\mathrm{gas},
\end{equation}
where $c^\star$ is a star formation efficiency factor:
\begin{equation}
c^\star = \mathrm{min}\left(0.2\frac{\Delta t}{t_\mathrm{dyn}},0.9\right).
\end{equation}
Here, $\Delta t$ is the length of the gas cell's timestep and $t_\mathrm{dyn}$ is the dynamical time of the gas cell (although note that a minimum value of $t_\mathrm{dyn}$=1\,Myr is imposed). The star particle also inherits the velocity and metallicity of its parent gas cell.  

\subsection{Stellar Feedback}\label{sec:feedback}

Stellar feedback in the form of Type II supernovae (SNe) is implemented in the FOGGIE simulations following \citet{Cen1992} and \citet{Cen2006}, with distributed feedback modifications from \citet{Smith2011}. Over the course of 12 dynamical times following its formation, a star particle injects a total of $10^{-5}\times M_\mathrm{\star,part}c^2$ thermal energy into the surrounding 27 gas cells. Note that we do not attempt to account for more delayed feedback (e.g., Type Ia SNe). During each timestep while $t<12\,t_\mathrm{dyn}$, the star particle also returns a fraction of its mass to the nearest 27 gas cells:
\begin{equation}
M_\mathrm{ret} = M_{\star,0}[(1+x_1)e^{-x_1}-(1+x_2)e^{-x_2}],
\end{equation}
where $M_{\star,0}$ is the initial mass of the star particle, 
\begin{equation}
x_1 = \frac{t-t_0}{t_\mathrm{dyn}},
\end{equation}
\begin{equation}
x_2 = \frac{t+\Delta t-t_0}{t_\mathrm{dyn}},
\end{equation}
and $t_0$ is the time at which the star particle was formed. By the end of this period (typically $\lesssim$100\,Myr) the star particle will have lost 25\% of its initial mass. The mass of metals returned to the gas, accounting for the recycling of gas back into stars, is
\begin{equation}
M_\mathrm{met} = 0.025\,M_{\star,0}(1-Z_\star)+0.25\,Z_\star,
\end{equation}
where $Z_\star$ is the metallicity of the star particle.

\subsection{Subhalo Finding and Merger Trees}\label{sec:halofinding}

Snapshots from the FOGGIE simulations are saved at a cadence of 5.4\,Myr. In each snapshot beginning at $z\approx6$, dark matter halos within the zoom region are identified with the \textsc{ROCKSTAR} halo finder \citep{Behroozi2013a}, which uses a friends-of-friends algorithm in combination with temporal and 6D phase-space information. Virial quantities $R_\mathrm{vir}$ and $M_\mathrm{vir}$ for each halo are also calculated by \textsc{ROCKSTAR} using the redshift-dependent $\rho_\mathrm{vir}$ of \citet{Bryan1998}. Table \ref{tab:simprops} lists the basic properties of each of the central FOGGIE galaxies in ascending order of $M_\star$. Unless otherwise specified, halo properties (e.g., $M_\star$) throughout this paper are based on all particles within $R_\mathrm{vir}$. Merger histories for each halo are assembled with \textsc{Consistent-Trees} \citep{Behroozi2013b} and halo properties are collated across time using \textsc{tangos} \citep{pontzen2018}.
\begin{table}
\centering
\caption{Properties of central FOGGIE galaxies at $z=0$}
\begin{tabular}{ccccc}
\hline
\hline
Name & $R_\mathrm{vir}$\footnote{Note that these values differ slightly from those listed in Table 1 of \citet{Lochhaas2021}, which uses $R_\mathrm{200}$, $M_\mathrm{200}$, etc.} & $M_\mathrm{vir}$ & $M_\star$  & $M_\mathrm{SH}$ \\
 & [kpc] & [10$^{12}$ M$_\odot$] & [10$^{10}$ M$_\odot$] & [10$^{10}$ M$_\odot$]\\
\hline
Tempest & 201 & 0.45 & 5.44 & 0.32 \\
Maelstrom & 253 & 0.90 & 11.6 & 0.86 \\
Squall & 235 & 0.76 & 12.6 & 1.20 \\
Blizzard & 261 & 0.99 & 14.7 & 1.76 \\
Hurricane & 301 & 1.05 & 25.7 & 2.50 \\
\hline
\end{tabular}
\label{tab:simprops}
\end{table}

\subsection{Caveats}
\label{caveats}
Although the mass resolution of the star particles and the gas in the FOGGIE simulations is state-of-the-art for cosmological simulations, the subgrid routines---particularly those associated with star formation and feedback---are imperfect and we consider here how this may impact our findings. 

As noted in \S\,\ref{sec:feedback}, these simulations employ exclusively thermal feedback and do not attempt to account for sources of feedback other than Type II SNe. Because no sources of delayed feedback, such as Type Ia SNe, are included, feedback occurs only over a relatively short period of time ($\sim100$\,Myr) following the formation of a star particle. Additionally, each individual star particle injects less energy into the surrounding medium than it would in a simulation that incorporated more complex feedback routines. The FOGGIE simulations also do not include a prescription for AGN feedback. While this likely has little impact on the many dwarf galaxies that contribute to the stellar halo \citep[although see, e.g.,][]{Sharma2020}, the central FOGGIE galaxies occupy a mass regime where AGN feedback may play a role in regulating galaxy growth \citep[e.g.,][]{Shankar2006,Keller2016}. 

Underpowered feedback has been shown to result in overproduction of stars and runaway growth of bulges, as galaxies cannot eject enough low angular momentum gas to effectively regulate their star formation \citep[e.g.,][]{Maller2002,Ceverino2009,Governato2010}. In the FOGGIE simulations, this is likely also compounded by a recently discovered issue with the star formation recipe described in \S\,\ref{sec:sf} that causes star formation to be slightly overefficient in dense regions and slightly underefficient in more diffuse regions. As a result, the galaxies are forming too many star particles in regions where the gas is most enriched and then failing to eject enough of this metal-rich gas, further enriching future generations of stars. We can therefore also anticipate that this combination of issues will produce galaxies with above average metallicities \citep[e.g.,][]{Brook2004,Brooks2007}. We also note that metal yields remain uncertain \citep[e.g.,][]{Peeples2014,Weinberg2023} and the normalization of the simulated vs observed mass-metallicity relation is therefore also uncertain \citep[e.g.,][]{D'Souza2018}.

We will refer back to these caveats throughout the paper when they are relevant to the results being discussed. However, as we will show, the FOGGIE stellar halos are typically consistent with observations. Additionally, the simulations do not need to be perfect for the relative differences between the simulated stellar halos to yield insights about the factors that influence the structure and assembly of stellar halos or inform plans for future observational strategies. \\ \\

\section{Selection of Stellar Halos}
\label{kinid}

At $z=0$, all five of the central FOGGIE galaxies are bulge-dominated disks surrounded by a diffuse and extended halo of stars. In this section, we describe how we separate out the star particles that populate the stellar halo from those that belong to the bulge or disk.

In order to identify the star particles that make up the stellar halo, we use a combination of kinematic information and position. We first identify the plane of the disk for each galaxy. In three of our five galaxies (Tempest, Maelstrom, and Squall), the disk is defined by the orbital angular momentum vector of young stars (age $<10$\,Myr) in the inner 15\,kpc of the galaxy. Blizzard is in the process of rejuvenating following a period of quiescence and therefore has very few extremely young stars, so its stellar disk is defined by the orbital angular momentum vector of stars with age $< 500$\,Myr in its inner 15\,kpc. The final galaxy, Hurricane, is a polar ring galaxy at $z=0$, so we identify two distinct disks: a central disk and a polar disk. Because both disks contain fairly young stars, but are tilted $\sim$80$^\circ$ with respect to one another, we cannot effectively use the orbital angular momentum of the young stars to identify the plane of either disk. Instead, we use the orbital angular momentum of cold gas ($T<10^4$\,K) within the inner 7\,kpc to identify the plane of the central disk and the orbital angular momentum of all gas with $r = 7$--25\,kpc to identify the plane of the polar disk. 

For each galaxy, we then place the disk in the x-y plane and calculate the orbital circularity for each star particle following \citet{Stinson2010}:
\begin{equation}
\epsilon = j_\mathrm{z}/j_\mathrm{circ},
\end{equation}
where $j_\mathrm{z}$ is the specific angular momentum of the star particle within the plane of the disk and $j_\mathrm{circ}$ is the specific angular momentum of a star in an ideal circular orbit located at the same radius within the plane of the disk. That is,
\begin{equation}
j_\mathrm{circ} = v_\mathrm{circ}r_\mathrm{xy}, 
\end{equation}
where $r_\mathrm{xy}$ is the distance of the star particle from the center of the galaxy within the x-y plane and $v_\mathrm{circ}$ is the circular velocity of a star orbiting at this radius. Note that this is distinct from the more commonly-used orbital circularity parameter defined by \citet{Abadi2003}, which defines $j_\mathrm{circ}$ as the specific angular momentum of a star in a circular orbit with the same binding energy as the star particle in question \citep[used by, e.g.,][]{Zolotov2009,Font2011,Cooper2015,Monachesi2016sims}. We choose to use the \citet{Stinson2010} parameter for the sake of computational efficiency. However, the use of one definition as opposed to the other has only a minor impact on the assignments of stars within the inner halo (and no impact on stars with $r>r_\mathrm{disk}$). Assuming that similar radial cuts are made for the disk and bulge components, the \citet{Stinson2010} kinematic cuts made here and the most commonly adopted \citet{Abadi2003} kinematic cut ($\epsilon<$0.8; see \citet{Monachesi2016sims} for a discussion of various $\epsilon$ cuts) identify 85--95\% of the same stars as belonging to the stellar halo. 

We expect star particles belonging to the disk to have $j_\mathrm{z} \approx j_\mathrm{circ}$, so we consider any star particle with $\epsilon = 0.65-1.3$ and $r<30$ kpc to be a disk star. Following \citet{Stinson2010}, \citet{Cooper2015}, and \citet{Monachesi2019}, star particles with non-disk orbits (i.e., with any value of $\epsilon > 1.3$ or $<0.65$) located within 5\,kpc of the center of the galaxy are considered to be members of the bulge. All other star particles within 350\,kpc of each galaxy at $z=0$ are classified as members of the stellar halo. We choose to limit our analysis to star particles within 350\,kpc because this radius comfortably encloses the virial radii of all of our galaxies (see Table \ref{tab:simprops}) without intersecting the stellar halos of any nearby massive neighbors. However, we note that all of our galaxies (and particularly the more massive galaxies, Blizzard and Hurricane) have star particles at larger radii that could reasonably be included in the stellar halo. As there is very little mass in the outskirts of the halo, including these distant star particles would have very little effect on any of our findings. 

It is worth noting that some authors choose to include only material that has been accreted from other galaxies in their stellar halo analyses \citep[e.g.,][]{Cooper2010,Deason2016,D'Souza2018}. However, because in situ and ex situ material are not easily distinguished in most observations and their relative fractions are therefore not well constrained, we do not separate out the accreted components of our stellar halos in the bulk of our analysis. We address the relative contributions of in situ and ex situ stars in \S\,\ref{dis}.

\begin{figure}
\centering
\includegraphics[width=0.47\textwidth]{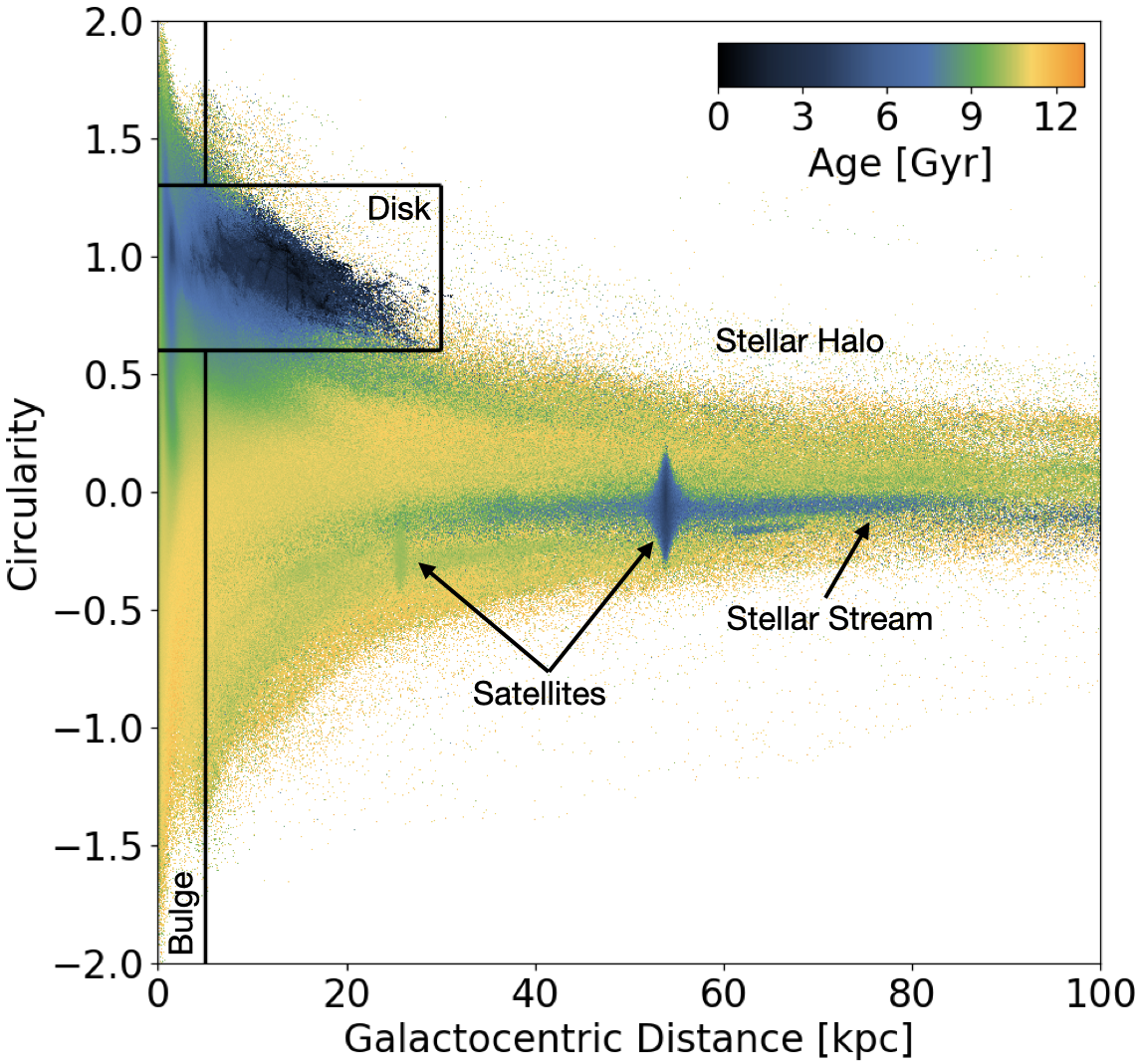}
\caption{Galactocentric distance and orbital circularity ($j_\mathrm{z}/j_\mathrm{circ}$) for all stars within 100\,kpc of the center of FOGGIE galaxy Tempest. Each star is colored according to its age at $z=0$. Stars are classified as part of the disk, the bulge, or the halo according to their location in the distance-circularity plane. As we might expect, the majority of the stars that make up the disk are relatively young, while those that compose the bulge and the stellar halo tend to be older. Satellites and stellar streams are also apparent as distinct substructures within the stellar halo.}
\label{fig:kinid}
\end{figure}

We show how our adopted definition separates out different components of a galaxy in Figure \ref{fig:kinid}. Each star particle within 100\,kpc of the center of Tempest is plotted in the radius-circularity plane and colored by its age at $z=0$. While the circularity varies from $\epsilon\,\approx -2$ to 2, the majority of the star particles with $\epsilon\,\approx 1$ (disk stars) are relatively young. By contrast, the bulk of the star particles that make up the bulge and the stellar halo are $\geq 10$\,Gyr old. Two satellites with intact cores are also apparent as distinct substructures within the stellar halo. The star particles bound to each occupy a fairly broad range of circularities due to the satellites' internal motions, but only a narrow range of radii, so they appear as vertical bands composed of somewhat younger stars than those that make up the majority of the halo. Both satellites are also in the process of being disrupted by Tempest's tidal forces, so their recently stripped star particles form nearly horizontal bands in the radius-circularity plane as they spread out along their original hosts' orbits.

In order to better mimic observations, which typically mask out the cores of bright satellites when analyzing stellar halos \citep[e.g.,][]{Gilbert2009,Jang2020,Gilhuly2022}, we will generally exclude those star particles still associated with satellites from our analysis. We find that masking out all star particles within the radius at which the g-band surface brightness of the satellite drops below 31.5 mag\,arcsec$^{-2}$ allows us to remove the core of the satellite without eliminating extended envelopes or other tidal features that are generally considered to be part of the stellar halo. The resulting stellar halo masses are listed in the final column of Table \ref{tab:simprops}. Note that we will also use alternative, observationally-motivated definitions for the stellar halo when appropriate (see Section \ref{mass}). \\

\section{Properties of FOGGIE Stellar Halos}
\label{props}
In this section, we summarize the properties of the FOGGIE stellar halos at $z=0$ and compare them to observed stellar halos. In \S\,\ref{mass}, we examine how different stellar halo definitions influence the mass of stellar halos. We present surface brightness maps and profiles of the stellar halos in \S\,\ref{sb}, with a description of the metallicity and color gradients of the halos in \S\,\ref{metcol}.

\subsection{Mass}
\label{mass}

\begin{figure}
\centering
\includegraphics[trim= 0mm 0mm 0mm 0mm, clip, width=0.47\textwidth]{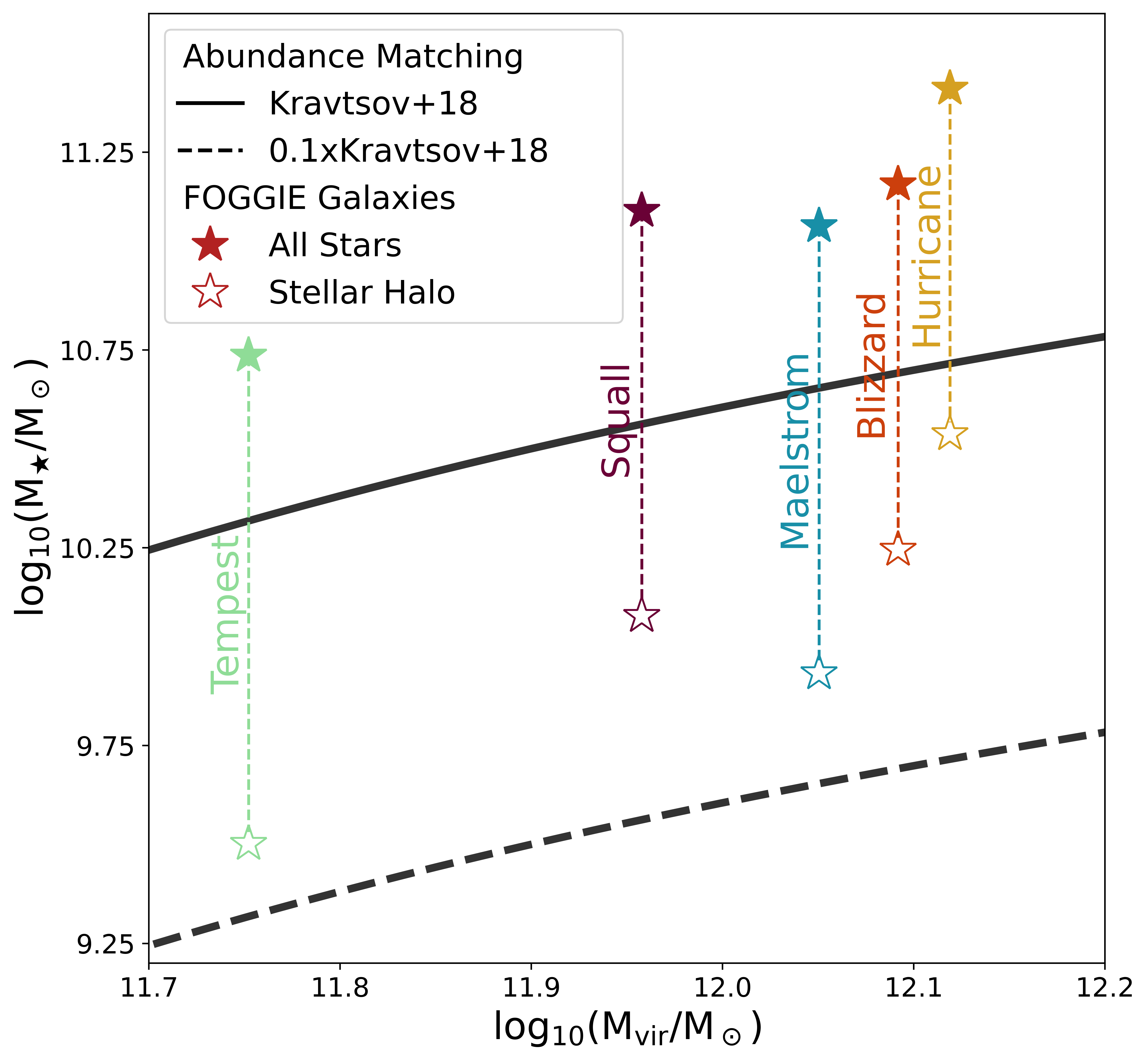}
\caption{FOGGIE stellar masses (solid stars) and stellar halo masses (outlined stars) compared to abundance matching results from \citet{Kravtsov2018}. Note that the virial masses from the simulations are adjusted following \citet{Munshi2013} and therefore differ slightly from those listed in Table \ref{tab:simprops}. Vertical dashed lines connect stellar masses and stellar halo masses for individual galaxies, the names of which are written alongside the vertical lines. We will use the same colors to refer to the same galaxies throughout this paper. The black dashed line shows 10\% of the predicted stellar mass for a given virial mass assuming \citet{Kravtsov2018} values. Each FOGGIE stellar halo makes up 6-13\% of its galaxy's total stellar mass.}
\label{fig:SMHM}
\end{figure}

In Figure \ref{fig:SMHM}, we show the stellar masses, virial masses, and stellar halo masses of the five FOGGIE galaxies compared to abundance matching results from \citet{Kravtsov2018}. In order to make a direct comparison to \citet{Kravtsov2018}, we multiply the virial mass of each galaxy by a factor of 1.25 to compensate for mass loss due to baryonic processes (e.g., supernova-driven gas outflows), following \citet{Munshi2013}.

As we noted in \S\,\ref{caveats} and as is evident in this figure, the FOGGIE galaxies have higher total stellar masses than we would expect for galaxies of their virial mass. They occupy dark matter halos spanning less than 0.5 dex in mass, with only slightly more scatter ($\sim0.7$\,dex) in total stellar mass. Their stellar halos range over nearly 1\,dex in mass---from $\sim$3$\times$10$^9$ M$_\odot$ \citep[similar to recent estimates of the mass of the Milky Way's anemic stellar halo---see, e.g.,][]{Deason2019,Mackereth2020} to $\sim$3$\times$10$^{10}$ M$_\odot$ (more akin to the rich stellar halo of M31 as measured by, e.g., \citet[][]{Ibata2014}). In agreement with both observations \citep[e.g.,][]{D'Souza2014} and previous simulations \citep[e.g.,][]{Purcell2007,Elias2018}, the stellar halo masses of the FOGGIE galaxies are correlated with their stellar and virial masses. 

Observations find that stellar halos typically make up 0.2--14\% of a galaxy's total stellar mass \citep[e.g.,][]{Harmsen2017,Jang2020,Gilhuly2022}. We include a dashed line corresponding to 10\% of the predicted \citet{Kravtsov2018} stellar mass in Figure \ref{fig:SMHM} to guide the eye of the reader in gauging the fraction of mass contributed by each stellar halo. The FOGGIE stellar halos make up between 6\% (Tempest) and 13\% (Hurricane) of their galaxy's total stellar mass. This is broadly consistent with observations, although we do not have any stellar halos at the extreme low end of the observed range.

\begin{figure*}
\centering
\includegraphics[trim= 0mm 0mm 0mm 0mm, clip, width=0.8\textwidth]{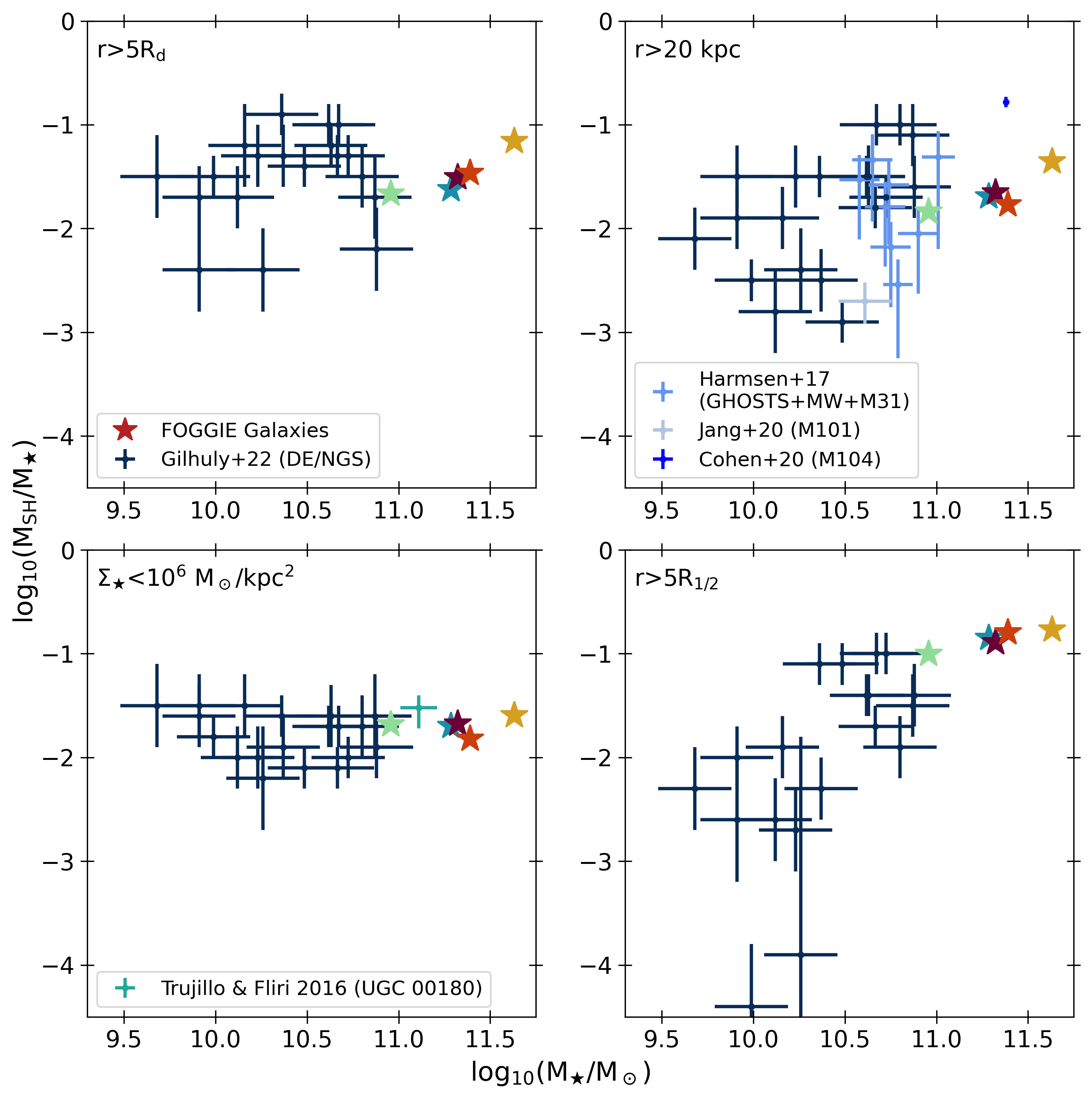}
\caption{FOGGIE stellar halo mass fractions compared to observations that adopt different definitions of the stellar halo. In each panel, the FOGGIE values are recalculated according to each observational definition: \textit{Top left:} all stellar mass beyond 5 times the scale radius of the disk; \textit{Top right:} all stellar mass beyond 20 kpc; \textit{Bottom left:} all stellar mass beyond the radius at which the stellar mass density drops below $10^6$ M$_\odot$\,kpc$^{-2}$; \textit{Bottom right:} all stellar mass beyond 5 times the stellar half mass radius. The FOGGIE stellar halos are generally consistent with observations. Compare to Figure 11 of \citet{Gilhuly2022}.}
\label{fig:masscomp}
\end{figure*}

There is no consensus method to define a stellar halo for either observations or simulations, which complicates any comparisons made between them \citep[e.g.,][]{Sanderson2018,Merritt2020,Gilhuly2022}. We make a more careful comparison to observations in Figure \ref{fig:masscomp} by reproducing Figure 11 of \citet{Gilhuly2022} with the FOGGIE galaxies. In order to account for uncertainty in the proper definition, we follow \citet{Gilhuly2022} and adopt four different commonly used definitions for the stellar halo. While disk stars and bulge stars are removed via kinematic information in our fiducial definition, we make only position-based cuts for these comparisons to better mimic observations. Note, however, that we still remove star particles bound to satellites from the calculation of the stellar halo masses, as observations typically mask these out.  

In each panel of Figure \ref{fig:masscomp}, the $x$-axis is the total stellar mass and the $y$-axis is the fraction of this stellar mass that is identified as part of the stellar halo by each observational definition. We include both the FOGGIE galaxies and samples of observed galaxies, which are plotted in the panel that most closely resembles each observational definition of a stellar halo. Note that the FOGGIE galaxies (with the exception of Tempest) are generally slightly more massive than most of the observations, although where possible we include data from M104 \citep{Cohen2020,Karachentsev2020} and UGC 00180 \citep{Trujillo2016}, which are similar in mass to Maelstrom, Squall, and Blizzard.

In the top left panel of Figure \ref{fig:masscomp}, the stellar halo is defined as the stellar mass located beyond 5 times the scale radius of the disk ($R_\mathrm{d}$). We calculate $R_\mathrm{d}$ by fitting an exponential profile to the face-on g-band surface brightness profile of each galaxy, excluding the bulge. The equation for an exponential fit is given by 
\begin{equation}
    \mu(r) = \mu_0+1.086\frac{r}{R_d},
\end{equation}
where $\mu_0$ is the central surface brightness of the galaxy in mag\,arcsec$^{-2}$ \citep{freeman1970}. The disk scale radius ranges from $R_\mathrm{d} = 2.7$--3.5\,kpc for the FOGGIE galaxies, which is consistent with the R$_\mathrm{d}$=2--3.5 kpc estimated for the Milky Way \citep{Bovy2012}. We compare the resulting stellar halo mass fractions to those found for the Dragonfly Edge-on Galaxies Survey and the Dragonfly Nearby Galaxies Survey \citep[which we will combine as DE/NGS;][]{Gilhuly2022}, both of which are integrated light surveys of the outskirts of local ($D<25$\,Mpc) spiral galaxies. DEGS \citep{Gilhuly2022} analyzed the stellar halos of 12 edge-on galaxies with masses greater than that of the LMC, while DNGS \citep{Merritt2016} studied the stellar halos of 8 Milky Way analogs without regard to orientation. Stellar masses for the DE/NGS galaxies come from the \textit{Spitzer} Survey of Stellar Structure in Galaxies \citep[S$^4$G;][]{Querejeta2015}, while the stellar halo mass fractions come from \citet{Gilhuly2022}. The FOGGIE galaxies are generally consistent with the observations, although they do not have as much diversity in either stellar mass or stellar halo mass fraction as the DE/NGS sample. 

In the top right panel of Figure \ref{fig:masscomp}, the stellar halo is defined as the stellar mass located beyond 20 kpc. In addition to the DE/NGS galaxies, we compare the FOGGIE galaxies to M101 \citep{Jang2020}, M104 \citep{Cohen2020}, and the galaxies of the GHOSTS survey \citep{Radburn-Smith2011,Harmsen2017}, plus the values \citet{Harmsen2017} adopted for the Milky Way \citep{Licquia2015,Bland-Hawthorn2016} and M31 \citep{Sick2015,Ibata2014}. The GHOSTS sample includes 6 nearly edge-on galaxies with stellar masses similar to that of the Milky Way. Following \citet{Gilhuly2022}, we use the uncorrected stellar halo masses for the GHOSTS galaxies and M104. The FOGGIE galaxies are again consistent with the observations and display a slight trend in stellar halo mass fraction as stellar mass increases.

In the lower left panel of Figure \ref{fig:masscomp}, the stellar halo is defined as the stellar mass located beyond the point at which the stellar surface density drops below 10$^6$\,M$_\odot$\,kpc$^{-2}$. This radius ranges from 15-30 kpc for the FOGGIE galaxies. We compare our galaxies to the DE/NGS galaxies as well as UGC 00180 \citep{Trujillo2016}. This definition of the stellar halo produces a stellar halo mass fraction that is nearly constant across 2 dex in stellar mass in both the observations and the FOGGIE galaxies.

In the lower right panel of Figure \ref{fig:masscomp}, the stellar halo is defined as the stellar mass located beyond 5 times the stellar half mass radius ($R_\mathrm{1/2}$) of each galaxy. The FOGGIE galaxies are within the scatter of the DE/NGS sample, but they sit at the very top of this range. As we discussed in \S\,\ref{caveats}, the bulges in the central FOGGIE galaxies tend to be slightly overmassive. The stellar half-mass radius of each galaxy is therefore only $\sim$1 kpc---much smaller than it otherwise would be (and smaller than it typically is for observed galaxies). Accordingly, $r>5 R_\mathrm{1/2}$ includes the bulk of the stellar disk mass in the calculation of the stellar halo mass. 

\citet{Harmsen2017} find that current simulations, regardless of resolution, tend to produce overly massive stellar halos---a trend that is also supported by the findings of \cite{Merritt2020} and \cite{Gilhuly2022}. The FOGGIE stellar halos are towards the high mass end of the observations when stellar halo definitions that rely on a measurement that includes the galaxy's bulge are used. Outside of this, however, the FOGGIE simulations appear to produce stellar halos with realistic masses. We discuss possible explanations for this in \S\,\ref{ishalo}.

\subsection{Surface Brightness}
\label{sb}
\subsubsection{Simulated Surface Brightness Limits}
\label{sblim}

Although surface brightness limits are frequently taken into account when the completeness of a survey is being calculated, it is less commonly appreciated that simulations are also limited in modeling low stellar densities by their discretization of stellar populations with star particles. \citet{Canas2020}, for instance, show that the vast majority of Milky Way-mass systems in \textsc{Horizon-AGN} have fewer than 100 star particles in what they call the ``intra-halo stellar component''. Assuming a stellar halo that fills the volume between 20 and 200\,kpc, this implies an average of only one star particle per (70\,kpc)$^3$. Given the amount of substructure that has been observed in real stellar halos, it is unlikely that large volume simulations---or even many zoom-in simulations---will be able to model the complex components of stellar halos in a realistic way.

The star particle formation scheme used in FOGGIE, and which we described in \S\,\ref{sec:sf}, preferentially places star particles with lower masses in regions that primarily consist of old stars, like the stellar halo. Accordingly, our stellar halos are made up of large numbers of particles: each comprises $2.9\times$10$^6$--$1.7\times$10$^7$ star particles (including particles bound to satellites increases these counts by $\lesssim$ 10\%). At $z=0$, the median star particle mass in the FOGGIE stellar halos ranges from 919\,M$_\odot$ (Tempest) to 1216\,M$_\odot$ (Squall)---similar to the mass resolution achieved in the highest-resolution Milky Way zoom-in simulations at the time of writing: \textsc{NIHAO}'s UHD simulations \citep{Buck2020}, \textsc{ChaNGa}'s Mint Condition DC Justice League simulations \citep{Applebaum2021}, and \textsc{Auriga}'s Level 2 simulation \citep{Grand2021}.  

\begin{figure*}
\centering
\includegraphics[width=0.97\textwidth]{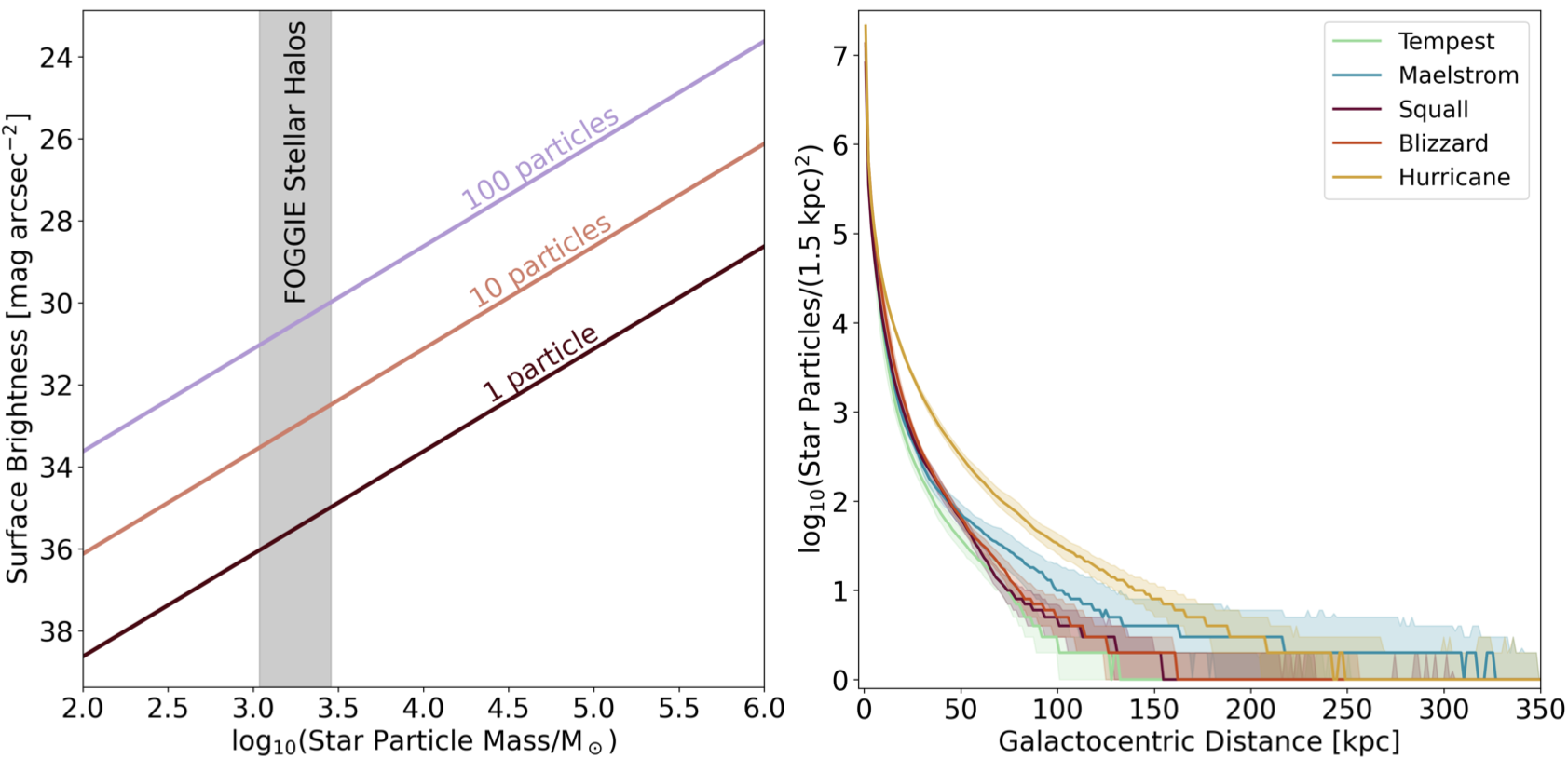}
\caption{\textit{Left:} Formation mass of star particles used in a simulation and the g-band surface brightness that these particles produce. Lines of different colors show the surface brightness of a single (1.5\,kpc)$^2$ area containing 1 (dark red), 10 (peach), or 100 (lavender) star particles. The dark gray vertical band indicates the interquartile range for star particle masses (at formation) in the stellar halos of the FOGGIE galaxies. \textit{Right:} The number of star particles contained within a (1.5\,kpc)$^2$ area as a function of radius in the various FOGGIE galaxies. The solid lines indicate the median value, while the shaded regions indicate the interquartile range \textit{assuming a non-zero detection}. While the disks of the FOGGIE galaxies typically contain $10^{4-8}$ star particles per unit area, the stellar halos typically only contain 1--10 particles per unit area. Lower mass star particles allow simulations to model lower surface brightness regions.}
\label{fig:mpsb}
\end{figure*}

We can convert star particle masses to approximate surface brightness limits by calculating the luminosity of each star particle and choosing a ``pixel'' area over which to measure the surface brightness. The latter is somewhat arbitrary, but effectively provides a normalization for the relationship between star particle masses and surface brightness. We use (1.5\,kpc)$^2$ areas throughout this paper because this is roughly the size of (10\,arcsec)$^2$---an area frequently used for standard surface brightness measurements \citep{Roman2020}---at the distance of M81. To calculate the luminosity of a star particle of a given mass, we use \textsc{FSPS} \citep{Conroy2009,Conroy2010} with MIST models \citep{Dotter2016,Choi2016,Paxton2011,Paxton2013,Paxton2015} and a \citet{Kroupa2001} IMF. We show the $g$-band surface brightness produced by a typical stellar halo star particle with age\,$= 12$\,Gyr and metallicity [Fe/H]\;$= -1.2$ as a dark red line in the left-hand panel of Figure \ref{fig:mpsb}. Because star particles are discrete, this line indicates the lowest surface brightness that a simulation using a given star particle mass can ``detect''. We show the interquartile range for the star particle masses (at formation time) in the FOGGIE stellar halos as a gray vertical band. The location of the intersection between the low end of this band ($\sim$10$^3$\,M$_\odot$) and the 1 particle/(1.5\,kpc)$^2$ line indicates that the detection limit of the FOGGIE simulations is $\sim$37 mag\,arcsec$^{-2}$.

In the right-hand panel of Figure \ref{fig:mpsb}, we show the median number of star particles contained within a (1.5\,kpc)$^2$ area as a function of galactocentric distance in the FOGGIE simulations. This value typically exceeds 10$^4$ star particles within the disk. However, those ``pixels'' within the stellar halo that contain star particles typically include only 10--100 star particles (shown as peach and lavender lines, respectively, in the left-hand plot) in the inner halo and 1--10 in the outer halo. While the low densities of star particles in outer halos remains a limitation, the FOGGIE galaxies and other simulations using M$_{\star\mathrm{,part}}<$1000\,M$_\odot$ offer a substantial improvement over previous generations of simulations. As we will show in \S\,\ref{SBmaps}, these simulations are resolving the surface brightness limits that will be most relevant for the wide-field surveys of the 2020s.

\subsubsection{Surface Brightness Maps and Profiles}
\label{SBmaps}
\begin{figure*}
\centering
\includegraphics[trim= 1mm 1mm 2mm 1mm, clip, width=0.97\textwidth]{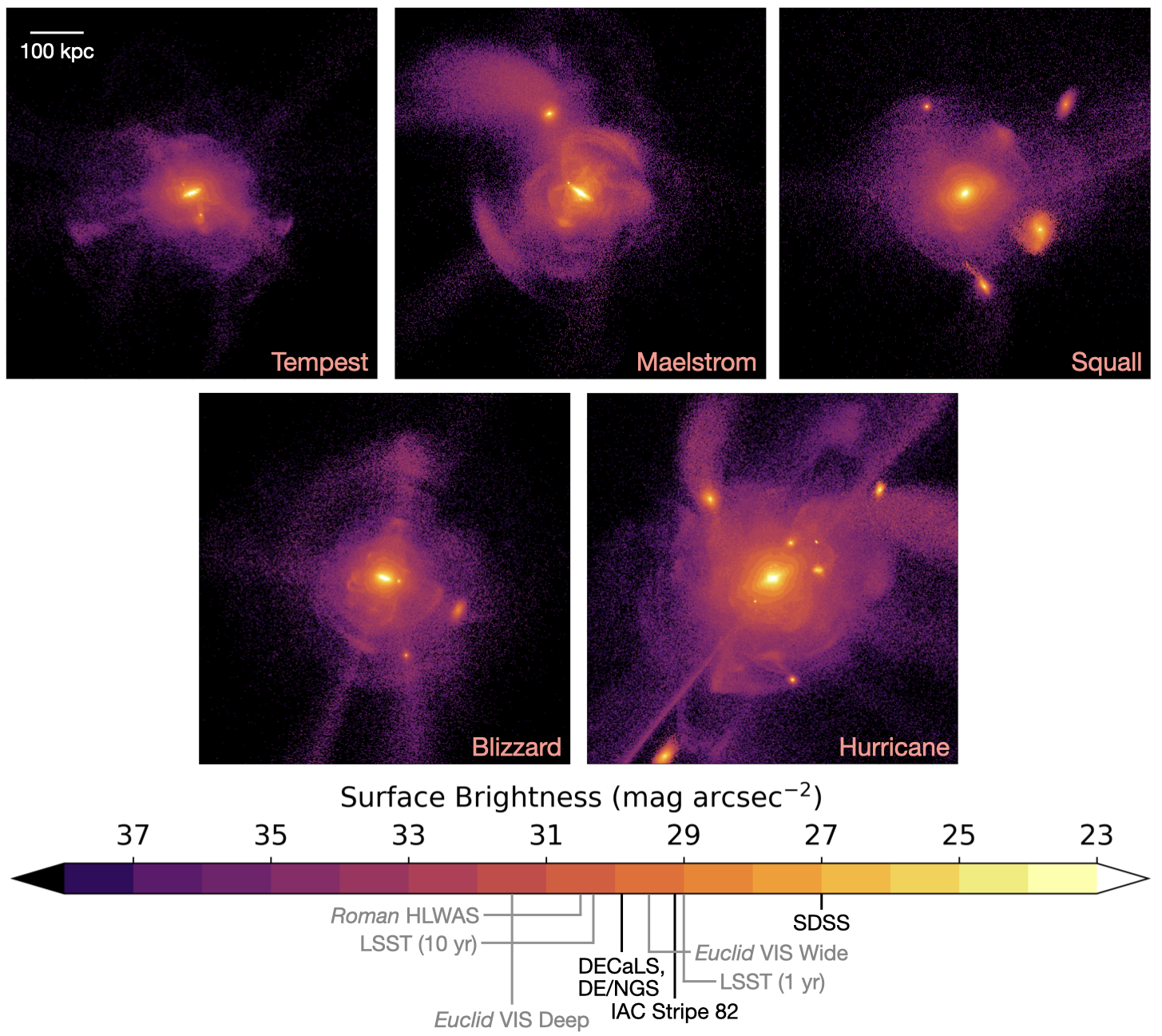}
\caption{Surface brightness maps of the five galaxies rendered in the $g$ band and oriented such that the stellar disks are edge-on. Each image is 700 kpc across and shows everything within 500 kpc of the center of each galaxy. The galaxies are ordered by increasing total stellar mass. The surface brightness ranges from 38 mag\,arcsec$^{-2}$ to 23 mag\,arcsec$^{-2}$ following Figures 13 and 14 of \citet{Bullock2005}. Labels below the colorbar mark 3$\sigma$ detection limits integrated over $10\times 10$\,arcsec$^2$ for SDSS \citep{Pohlen2006}, the Dragonfly Edge-on/Nearby Galaxies Surveys \citep[DE/NGS;][]{Merritt2016,Gilhuly2022}, IAC Stripe 82 \citep{Fliri2016,Roman2018}, Rubin LSST 1 and 10 year co-added data \citep{Yoachim2022}, the \textit{Euclid} VIS Wide and Deep Surveys \citep{EuclidCollaboration2022}, DECaLS \citep{Dey2019,Roman2021}, and the \textit{Roman} High Latitude Wide Area Survey \citep[HLWAS;][]{Martinez-Delgado2023,Montes2023}. Estimates for future surveys are shown in gray. Note that all limits except for those shown for \textit{Roman} and \textit{Euclid} are for $g$-band filters.}
\label{fig:haloim}
\end{figure*}

Despite spanning a narrow range of stellar and virial masses, the FOGGIE galaxies have diverse stellar halos. In Figure \ref{fig:haloim}, we show $g$-band surface brightness maps ranging from 38 mag\,arcsec$^{-2}$ to 23 mag\,arcsec$^{-2}$ for each of our 5 galaxies \citep[cf. Figures 13 \& 14 of][]{Bullock2005}. Note that we include all star particles falling within these projections out to a distance of 500\,kpc of the center of each galaxy, so that bulge stars, disk stars, and stars bound to satellites are included. Surface brightness is calculated as described in Section \ref{sblim} according to the mass, age, and metallicity of each star particle and using (1.5\,kpc)$^2$ pixels. Each galaxy is oriented such that its disk is edge-on (for Hurricane, we orient the image relative to its central disk). Below the surface brightness maps, we also show the approximate 10$\times$10 arcsec$^2$ surface brightness limits for a number of surveys (assuming 3$\sigma$ significance). We mark future surveys with uncertain final limits in gray. We use $g$-band (central wavelength\,$\approx$\,477 nm) limits wherever possible so that a direct comparison to the FOGGIE surface brightness maps can be made. However, for \textit{Euclid}, we use limits for the VIS instrument, which uses a single broad band filter that covers 550--900 nm. For \textit{Roman}, we use limits appropriate for filters F106 and F129, as these are the bluest filters that are currently planned for use in the High Latitude Wide Area Survey (HLWAS). For a more detailed discussion of these filters, see Section \ref{intlight}.

The FOGGIE galaxies are shown in order of increasing total stellar mass. While there is a broad trend of stellar halo extent and general complexity increasing with the stellar mass of the galaxy it surrounds, we have only a small sample of galaxies and there is certainly scatter arising from their diverse histories. Tempest has the most compact stellar halo, with only a few streams extending beyond 100\,kpc. Maelstrom's stellar halo is dominated by shell structures, but it also has a large stream resulting from the ongoing tidal disruption of an LMC-mass dwarf, the core of which is still visible in the upper left of the image. Squall's stellar halo is relatively smooth, with only a few shell structures evident. However, it has four prominent satellites that have fallen in recently and are just beginning to be tidally stripped. Blizzard has a number of extended streams---primarily stemming from the destruction of a single dwarf satellite---coupled with a series of shells. The most massive galaxy, Hurricane, also has the richest stellar halo. This is due in part to the large number of fairly bright satellites that have survived to $z=0$ with intact cores and extended tidal streams. However, Hurricane has also simply accreted and disrupted considerably more satellites than any of the other galaxies, leading to a more massive and complex halo. It is worth noting that the bulk of the variations evident in the outer halos of these galaxies lie at extremely low surface brightnesses ($\mu_g>$33.5 mag arcsec$^{-2}$) and would therefore likely be washed out in a simulation with M$_\star\mathrm{,part}\gtrsim$10$^4$\,M$_\odot$.

\begin{figure*}
\centering
\includegraphics[trim= 1mm 1mm 1mm 1mm, clip, width=0.97\textwidth]{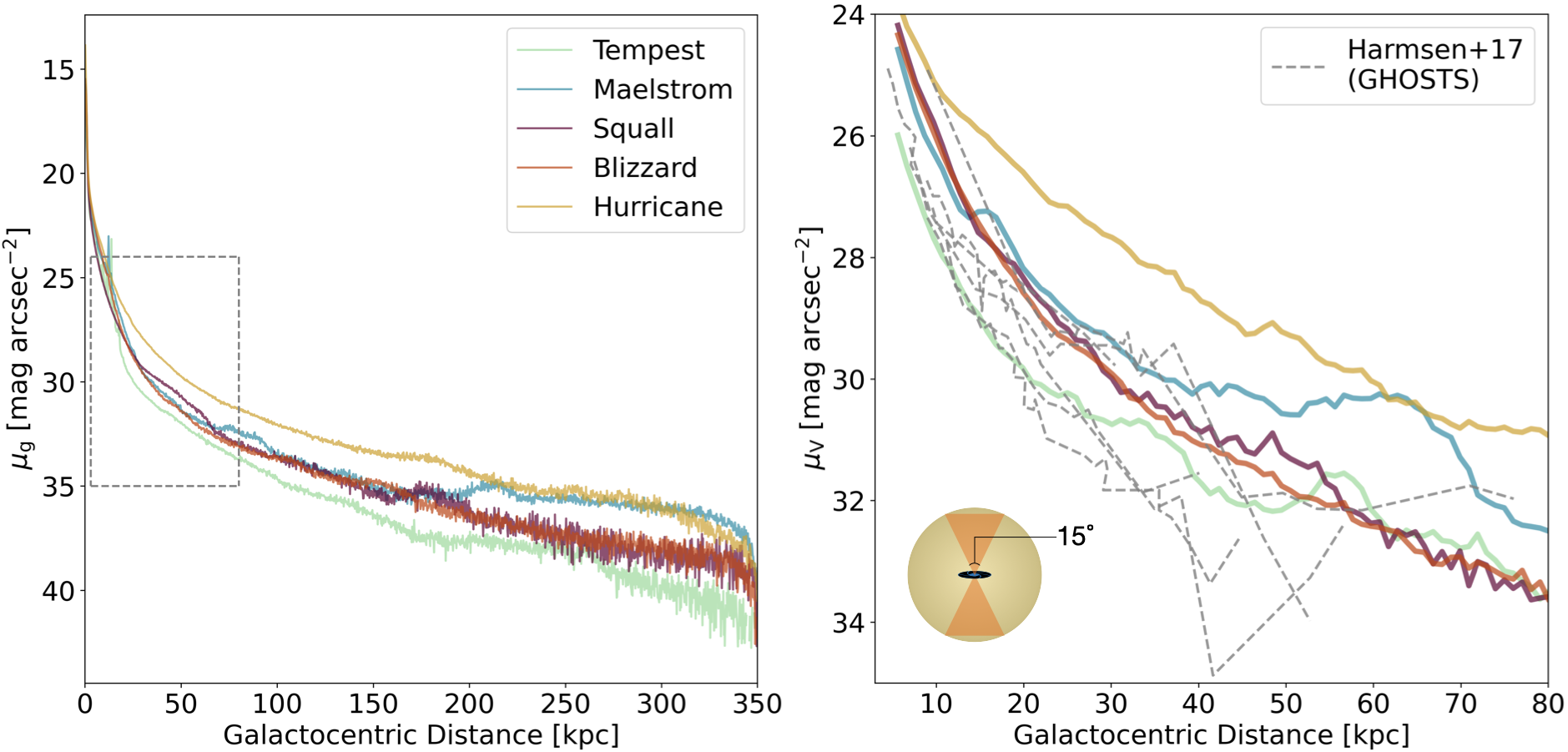}
\caption{\textit{Left:} Face-on g-band surface brightness profiles of the FOGGIE galaxies with star particles belonging to satellites excluded. The grey dotted rectangle indicates the range of the right-hand plot. \textit{Right:} V-band surface brightness profiles of the inner regions of the FOGGIE stellar halos. As shown in the cartoon in the bottom left, the profiles in this plot are measured only within a 15$^\circ$ wedge along the minor axis of the disk. This is done to provide a better comparison to the stellar halos in the GHOSTS survey \citep{Harmsen2017}, which are shown as dashed grey lines. The five FOGGIE halos typically span a range of $\sim$3 mag\,arcsec$^{-2}$ at any given radius, but, except the brightest halo, Hurricane, are broadly consistent with the GHOSTS galaxies.}
\label{fig:sbprof}
\end{figure*}

The differences between the FOGGIE stellar halos are also evident in Figure \ref{fig:sbprof}, where we show 1D surface brightness profiles of our galaxies. The overall shapes of the profiles are quite similar, but they vary significantly in brightness. In the left-hand panel, we show the azimuthally-averaged $g$-band surface brightness profile of each galaxy out to 350\,kpc. Note that stars still bound to satellites have been removed as described in Section \ref{kinid}. While Blizzard and Squall have nearly identical profiles, Hurricane is considerably brighter (1.5--2 mag\,arcsec$^{-2}$) than the other halos and Tempest is slightly dimmer (1--1.5 mag\,arcsec$^{-2}$). This is unsurprising given both the relative masses of the stellar halos (Figure \ref{fig:SMHM}) and their appearances (Figure \ref{fig:haloim}). Maelstrom's profile is nearly identical to those of Squall and Blizzard out to $\approx 200$\,kpc, but flattens out at this radius, rather than continuing to decline. This is due to the disruption of its LMC-mass satellite, the tidal debris from which dominates its outer halo.

In the right-hand panel of Figure \ref{fig:sbprof}, we compare the V-band surface brightness profiles of the FOGGIE galaxies to the stellar halos of the GHOSTS sample \citep{Harmsen2017}. The range covered by this plot is identified as a dotted rectangle in the left-hand panel of this figure. In order to make a direct comparison to the observations, we calculate this profile for only a subset of the stars. The GHOSTS profiles we compare to are measured along the minor axis of each observed galaxy between 5 and 40--75\,kpc in order to minimize contamination by disk stars. Following \citet{Monachesi2016sims}, we mimic this by orienting each FOGGIE galaxy such that the disk is edge-on and then select only those stars that fall within a projected 15 degree wedge above or below the disk, starting 5\,kpc from the center of the galaxy and excluding any stars bound to satellites (see the cartoon in the lower left-hand corner). The V-band profiles shown here include only these stars. Because Hurricane's polar ring is tilted at $\sim 80^\circ$ relative to its central disk, we rotate the minor axis wedges an extra 30$^\circ$ to avoid intersecting the polar ring. However, this has only a minor effect on the surface brightness profile that we derive.

All of the FOGGIE galaxies except Hurricane match the observations fairly well in shape and general normalization. As in Figure \ref{fig:sbprof}, Hurricane is considerably brighter than any of the other galaxies, including the observed ones. This is likely primarily due to the fact that Hurricane is the most massive of the galaxies and is more massive than any of the GHOSTS galaxies by $\sim 0.5$\,dex. Maelstrom also has a larger upturn in its surface brightness profile than we see in the observations starting at $\sim 40$\,kpc. This is the result of the intersection of its lower minor axis wedge with a stellar stream from a satellite with M$_\star\sim 10^9$\,M$_\odot$ at this radius. There is also a subtler flattening in the surface brightness profiles of the other three FOGGIE galaxies at $r\geq 40$\,kpc that may be inconsistent with the GHOSTS sample. Some of this can likely be attributed to the fact that Tempest is the only FOGGIE galaxy that is not above the mass range of the GHOSTS galaxies (see the upper right-hand panel of Figure \ref{fig:masscomp}). However, this slight excess of light may also be linked to broader inconsistencies that have been found between simulations and observations.

\citet{Merritt2020} carefully compared the stellar halos of the DNGS to a mass-matched sample from TNG100 and found that the simulated galaxies had stellar surface densities 1--2\,dex higher than the observed galaxies at $r > 20$\,kpc---a disparity that they dubbed the ``missing outskirts'' problem. \citet{Keller2022} found a potential explanation for this issue, showing that simulations run with feedback schemes that efficiently regulate star formation in high redshift and low mass halos produce stellar halos more in line with the observations than do simulations with more traditional feedback schemes. Given the simplicity of the feedback implemented in the FOGGIE simulations, it is perhaps not surprising that we see some evidence that our stellar halos have excess light. However, our simulated stellar halos are considerably more in line with both observed masses and observed surface brightness profiles than those studied in either \citet{Merritt2020} or \citet{Elias2018} (see Figure 11 of \citet{Gilhuly2022}). If our stellar halos do suffer from the missing outskirts problem, it is likely a relatively minor concern. We address a possible explanation for this in \S\,\ref{ishalo}.

\subsection{Metallicity and Color}
\label{metcol}
The stars that make up the stellar halos of Milky Way-like galaxies are generally observed to be relatively metal-poor \citep[e.g.,][]{Mouhcine2005,Harmsen2017}. However, considerable variation exists both between stellar halos and within them. The Milky Way and M31, for instance, appear to lie on opposite ends of the spectrum: the Milky Way's halo is diffuse and exceptionally metal-poor \citep[e.g.,][]{Bell2008}, while M31's halo is brighter and contains more metal-enriched stars \citep[e.g.,][]{Ibata2014}. The metallicities of other nearby galaxies of similar mass typically lie somewhere between the two \citep[e.g.,][]{Mouhcine2005,Harmsen2017}. Both observations \citep[e.g.,][]{Mouhcine2005} and simulations \citep[e.g.,][]{Renda2005,Robertson2005,Font2006,Purcell2008} suggest that these variations in metallicity are the result of differences in how the stellar halos are assembled. \citet{Deason2016} and \citet{D'Souza2018} show that the metallicity of the accreted component of stellar halos reflects the metallicity of their dominant contributor(s)---typically one to two dwarf galaxies with $M_\star \, = \, 10^{8-10} \, $M$_\odot$. The tight relationship that exists between the mass of a dwarf galaxy and its metallicity \citep[e.g.,][]{Gallazzi2005,Kirby2013} produces a similarly strong correlation between the mass of a stellar halo and its metallicity.

\begin{figure*}
\centering
\includegraphics[trim= 0mm 0mm 0mm 0mm, clip, width=0.97\textwidth]{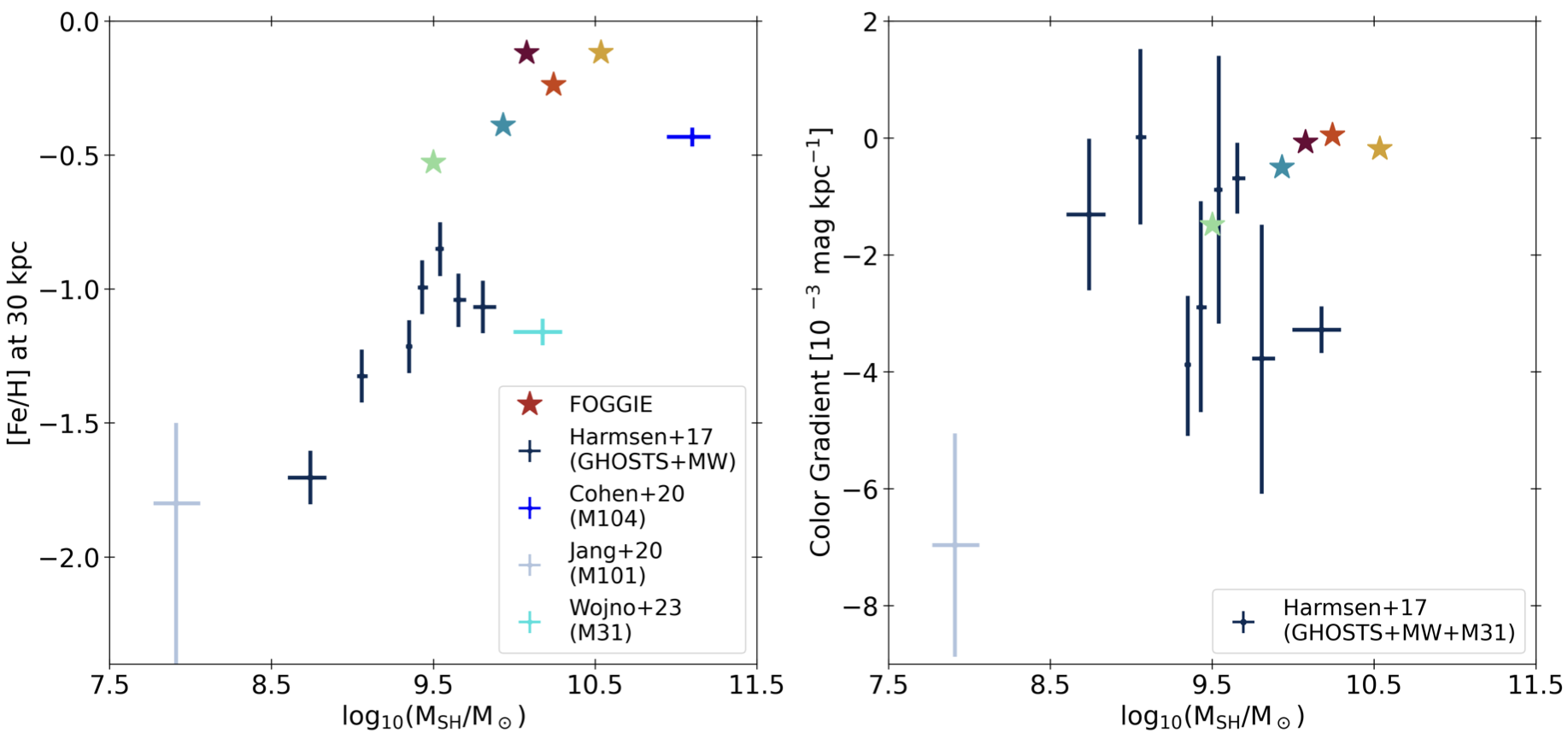}
\caption{\textit{Left:} The masses and metallicities of the FOGGIE stellar halos compared to observed samples. [Fe/H] is measured at 30 kpc from the center of each galaxy based on the stars that fall within a 15$^\circ$ wedge along the minor axis of the disk. Although the FOGGIE galaxies reproduce the slope of the observed relation, they are biased high by $\sim$0.4 dex. \textit{Right:} The masses of the FOGGIE stellar halos and the $F606W-F814W$ color gradient of RGB stars measured between 5 and 40 kpc along their minor axes, compared to observed samples. The FOGGIE galaxies typically have relatively flat color gradients, although Tempest has a slight negative color gradient.}
\label{fig:cmcomp}
\end{figure*}

In the left-hand panel of Figure \ref{fig:cmcomp}, we compare the masses and metallicities of the FOGGIE stellar halos to M101 \citep{Jang2020}, M104 \citep{Cohen2020}, and the galaxies in the GHOSTS sample \citep{Monachesi2016obs,Harmsen2017}. As in Figure \ref{fig:masscomp}, we also include the values used by \citet{Monachesi2016obs} and \citet{Harmsen2017} for the Milky Way, i.e., the mean of the metallicity values found by \citet{Sesar2011} and \citet{Xue2015}. However, for M31, we use a recent measurement from \citet{Wojno2023}, adopting their best-fit value for the smooth halo and substructure stars at 30 kpc (see their Figure 4). In order to make a direct comparison to the observations, we measure the metallicities of the FOGGIE stellar halos at $r = 30$\,kpc within the minor axis wedges defined in Section \ref{SBmaps}. Note that we do not track the abundances of individual elements in the FOGGIE simulations, so we have converted total metallicity to [Fe/H] following \citet{Thomas2003} and assuming [$\alpha$/Fe]\,=\,0.3, appropriate for an old, metal-poor population of stars \citep[e.g.,][]{Venn2004,Robertson2005}.

The observed galaxies show a strong correlation between stellar halo metallicity and mass, as do the FOGGIE galaxies. However, as we noted in \S\,\ref{caveats}, the FOGGIE galaxies are somewhat metal-rich with respect to the observations ($\sim 0.4$\,dex above the expected values). For this reason, we will largely abstain from commenting on absolute metallicities in the rest of our analysis. However, it is worth noting that the FOGGIE galaxies reproduce the slope of the observed $M_\mathrm{SH}$-metallicity relation, and it is therefore likely that relative differences between the FOGGIE stellar halos can be trusted. 

In the right-hand panel of Figure \ref{fig:cmcomp}, we explore how metallicity varies within the FOGGIE stellar halos. A combination of smaller galaxy sizes at high redshift and the mass-dependence of dynamical friction are thought to cause the inner halos of galaxies to be dominated by ancient and/or massive accretion events. The outskirts of the halos, by contrast, are thought to be primarily populated by stars from lower mass and/or more recent accretion events \citep[e.g.,][]{Bullock2005,Johnston2008,Horta2023}. We might, therefore, expect the dwarf galaxy mass-metallicity relation to produce a negative metallicity gradient in most stellar halos. However, the stochastic history of satellite accretion and the variations in decay time inherent in different satellite orbits lead to significant halo-to-halo variation. Flatter metallicity gradients generally indicate that many dwarfs have contributed fairly equally to a stellar halo, while a significant gradient is more indicative of a stellar halo dominated by one or two massive contributors \citep[e.g.,][]{Monachesi2016obs}.

In order to compare to observations of metallicity gradients in stellar halos, we follow \citet{Monachesi2016obs} in using the $F606W-F814W$ color gradient of red giant branch (RGB) stars as a proxy for metallicity. Although age and metallicity are degenerate \citep[e.g.,][]{Worthey1994}, metallicity has a stronger influence on the color of RGB stars than age does \citep[e.g.,][]{Streich2014}, and using color gradients, rather than metallicity gradients, avoids the large uncertainties (0.2--0.3 dex) associated with a conversion. We calculate the $F606W-F814W$ color of RGB stars in the FOGGIE simulations by subsampling each star particle that falls within a minor axis wedge assuming a \citet{Kroupa2001} IMF. We use MIST models to identify those ``stars'' that would be RGB stars and to calculate their $F606W$ and $F814W$ luminosities. The gradients plotted in Figure \ref{fig:cmcomp} are based on a linear fit to the $F606W-F814W$ color of these ``RGB stars'' between 5 and 40\,kpc.

The $F606W-F814W$ color gradients we measure are generally in good agreement with metallicity gradients over the same radial range. Squall, Blizzard, and Hurricane all have nearly flat gradients throughout most of their stellar halos, although we note that Hurricane has a negative metallicity gradient ($-0.008$\,dex/kpc) in its outer halo ($r=170$--220\,kpc), where debris from a low mass satellite passes through its minor axis. Tempest is the only galaxy with a substantial color gradient in its inner halo, although the change in the metallicity of its stellar halo within this region is very similar to that of Maelstrom (both are $-0.01$\,dex/kpc), which has a nearly flat color gradient. The presence of an inner halo gradient in both Tempest and Maelstrom likely reflects the fact that both have had relatively recent accretion events that left debris at low impact parameters.

The FOGGIE stellar halos are broadly in line with observations, although we do not have any stellar halos with extremely negative gradients. \citet{Monachesi2016obs} and \citet{Harmsen2017} find that roughly half of the GHOSTS stellar halos have no color/metallicity gradients, while the remainder have slightly negative gradients. In Figure \ref{fig:cmcomp}, we also show a $F606W-F814W$ gradient for M101 based on a linear fit to data from \citet{Jang2020}. M101 has a much lighter stellar halo with a much steeper negative gradient than we see in any of the FOGGIE galaxies. This is likely partially due to the fact that our simulated sample is small. However, this may also reflect the physical prescriptions used in the simulations. Models that result in stellar halos with high in situ fractions tend to also produce substantial negative metallicity gradients \citep[e.g.,][]{Font2011,Tissera2013}, while those that are exclusively accretion-based often produce no gradients \citep[e.g.,][]{Bullock2005,Font2006}. \citet{Monachesi2019} find that, although metallicity gradients are present even in the accreted components of many \textsc{Auriga} stellar halos, they are shallower than those found when the in situ stars are included. As we will discuss in Section \ref{ishalo}, the FOGGIE stellar halos generally have small contributions from in situ stars, particularly along their minor axes.

\section{(Dis)assembling Stellar Halos}
\label{dis}
Evidence from both theory and observations suggests that stellar halos are predominantly composed of stars that originally formed in other galaxies, outside of the main progenitor of the $z=0$ central \citep[e.g.,][]{Bullock2001,Bullock2005,Naidu2020}. By disassembling stellar halos into the various galaxies that contributed to them, we have the opportunity to learn about a wide variety of dwarf galaxies---not merely the small and biased subset that survived as gravitationally self-bound entities until the present day.

In this section, we will explore the origins of the star particles that contribute to the $z=0$ stellar halos of the FOGGIE galaxies. We describe the method by which we identify and classify the sources of stellar halo star particles in \S\,\ref{origins}, then discuss the in situ and ex situ star particles in \S\,\ref{ishalo} and \S\,\ref{acchalo}, respectively. Note that \S\,\ref{origins} is fairly technical, so those uninterested in the details of how stellar halo star particles are assigned to a given source should feel free to skim or skip it.

\subsection{Identifying Contributors to the Stellar Halo} \label{origins}
In order to trace each individual star particle in each $z=0$ stellar halo back to the (likely no longer gravitationally self-bound) galaxy in which it formed, we compare the location at which it formed to the locations of dark matter halos identified by \textsc{Rockstar} at that snapshot. The extremely fine time cadence of the FOGGIE simulations typically makes this fairly simple, as star particles are unlikely to move far from their birth sites in the $\sim 5$\,Myr that separate snapshots. If a star particle is within $0.2R_\mathrm{vir}$ of the center of a dark matter halo, we assume it formed within that dark matter halo. If a star particle is within $0.2R_\mathrm{vir}$ of the centers of multiple dark matter halos, we assume it formed within the one whose absolute distance to it is smallest. 

In instances where no appropriate host halo is found for a particular star particle, we employ several ``clean-up'' strategies. The first technique makes use of \textsc{Consistent-Tree}'s ``phantom'' halo feature, which identifies when \textsc{Rockstar} has temporarily lost track of a halo and interpolates the likely position and velocity of that halo during the time when it was lost. This is particularly common during mergers---including infall to a larger halo---which is incidentally one of the most critical times for galaxies contributing stars to a stellar halo. We use the estimated positions from \textsc{Consistent-Trees} to determine whether a hostless star particle was within $0.2R_\mathrm{vir}$ of the center of a phantom halo when it formed. If this procedure does not yield a host, we identify one manually. For each snapshot in which a hostless star particle forms, we plot the positions of the hostless star particle and any nearby star particles. In nearly all cases, the hostless star particle is obviously associated with a group of previously assigned star particles at the time that it forms and it is assigned to the same galaxy as its companions. The failure of the earlier procedures to identify hosts for these star particles is typically due to one of three issues: 1) the host halo has fallen below the limit where \textsc{Rockstar} is able to keep track of it, but is still forming stars; 2) the host halo is a phantom during the time when this star particle formed and \textsc{Consistent-Trees} did not correctly predict its position; 3) the host halo's dark matter halo has merged with that of another galaxy, but the stellar components of the galaxies are still separated by more than $0.2R_\mathrm{vir}$. In rare instances, we also see star formation occurring in tidal dwarfs or dark matter halos that were never massive enough for \textsc{Rockstar} to identify them. A star particle that formed in one of these locations cannot be assigned to a \textsc{Rockstar} halo, but it still receives the same host ID as any other star particles that formed in the same location.  

After each star particle that is part of a stellar halo at $z=0$ has been assigned a host ID corresponding to the galaxy that it originally formed in, we consolidate any hosts that merged prior to infall. We do this for two primary reasons: 1) If two galaxies have fully merged prior to infall to the central, their debris should occupy the same phase-space. Our results concerning contributors to the stellar halo are therefore more directly comparable to observations if we treat the two merged galaxies as a single entity. 2) The granularity of our initial host ID assignments are limited by our particle mass resolution. The mass of dark matter particles in the high resolution region of these simulations is $\sim 10^6$\,M$_\odot$, so we do not consider dark matter halos with a total mass $<10^9$\,M$_\odot$ ($\sim 1000$ dark matter particles) to be resolved. Accordingly, all analysis involving the decomposition of our stellar halos into their component parts should be assumed to lack ultra-faint dwarfs. The dwarf galaxies that contribute to our stellar halo are almost certainly composed of smaller dwarf building blocks and many should likely possess ultra-faint satellites of their own. However, simulations from \citet{Deason2016} suggest that stars from ultra-faint dwarfs make up $\ll$1\% of the mass of stellar halos around Milky Way-like galaxies. This dominance of more massive dwarfs is also supported by observations of the relative abundances of different populations of stars in the stellar halo \citep[e.g.,][]{Fiorentino2015,Deason2015}. By associating star particles with their host at the time at which they first fall into our central galaxy, we are limiting our analysis to galaxies that we can resolve without omitting significant contributors. As a final pass, we check that all star particles identified as belonging to a particular host halo make up only a single galaxy at infall (or prior to disruption if disruption precedes infall) and that no star particles have been incorrectly assigned to low-mass dark matter subhalos that happened to be closer to them than their true host at the time of formation. \\ \\

\subsection{The in situ halo}
\label{ishalo}
Although the majority of stars that populate the stellar halos of Milky Way-like galaxies are thought to have formed in disrupted dwarfs, there is evidence to suggest that some may have formed in situ---either within the disk of the main progenitor of the central galaxy or within the halo itself (i.e., in dense clumps within the CGM or in gas recently stripped from infalling dwarfs). Using SDSS, \citet{Carollo2007} found that the Milky Way appeared to have a more metal-rich inner halo ($r < 15$\,kpc) in addition to a metal-poor outer halo and theorized that these stars may have formed dissipatively, rather than arriving through accretion. More recent observations with \textit{Gaia} and the H3 Survey have also found a population of relatively metal-rich stars within the Milky Way's inner halo that are thought to have formed in situ \citep[e.g.,][]{Bonaca2017,Haywood2018,Conroy2019b}. 

Cosmological simulations typically predict that in situ stars contribute 20--50\% of the mass in stellar halos around Milky Way-mass galaxies and dominate the stellar halo mass budget out to 30-40\,kpc \cite[e.g.,][]{Abadi2006,Zolotov2009,McCarthy2012,Cooper2015,Font2020}. \citet{Cooper2015} find that in situ halo stars usually have one of three origins: some form as part of the central galaxy's disk and are displaced to larger radii via violent relaxation during a major merger \citep[see also][]{Zolotov2009,Purcell2010}, but others form either in gas that has been recently stripped from dwarf galaxies or in smoothly accreted gas. Another origin has been suggested by \citet{Yu2020}, who find that 5--40\% of the stars that populate the outer ($r>50$\,kpc) stellar halos of the galaxies in FIRE's \textsc{Latte} suite were formed in outflows from the central galaxy.

The FOGGIE stellar halos also include a contribution from in situ star particles. In line with other simulations, we find that 30--40\% of the mass in the stellar halos of Tempest, Maelstrom, Squall, and Hurricane and 58\% of the mass in Blizzard's stellar halo come from in situ star particles. We also find that the majority of these star particles either formed in the central disk of the main progenitor and were perturbed into halo orbits by mergers or formed in gas recently stripped from infalling dwarf galaxies. However, the in situ star particles in the FOGGIE stellar halos tend to be more centrally concentrated than those in other simulations. 

\begin{figure}
\centering
\includegraphics[trim= 0mm 0mm 0mm 0mm, clip, width=0.47\textwidth]{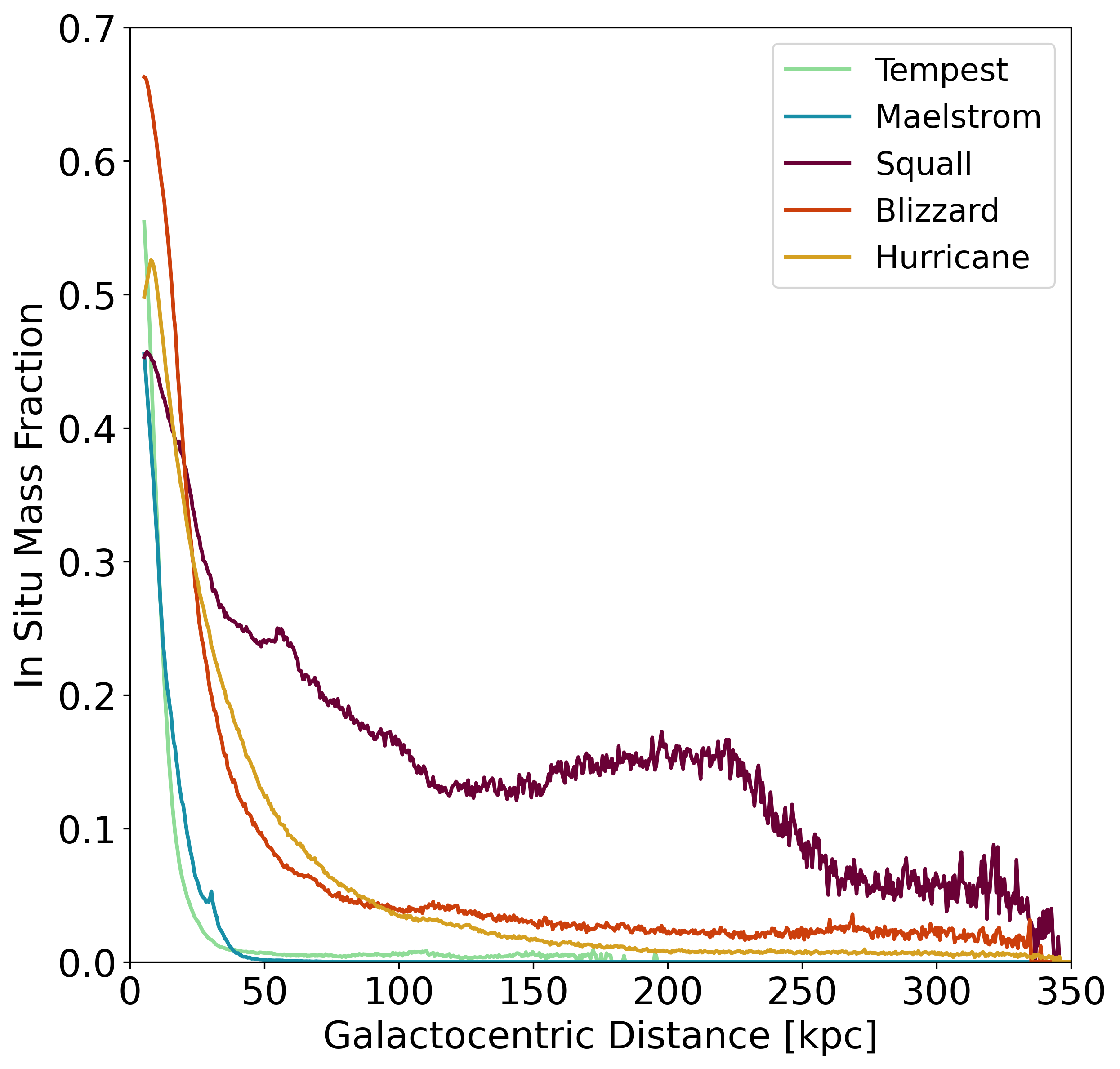}
\caption{The fraction of the stellar halo mass in a given annulus that comes from star particles that formed in the main progenitor of the central galaxy as a function of radius. The global in situ mass fractions of the FOGGIE stellar halos are similar to those found in other cosmological simulations, but the in situ populations of the FOGGIE stellar halos tend to be more centrally concentrated. The only exception is Squall, which experiences a relatively late major merger that perturbs a number of in situ star particles from its disk and inner halo into wider orbits.}
\label{fig:insitu}
\end{figure}

In Figure \ref{fig:insitu}, we show the in situ mass fraction of each stellar halo as a function of radius. We compute this by randomly orienting each galaxy 120 times and calculating the fraction of the mass within a given projected annulus (0.5 kpc in width) that is contributed by in situ star particles. The only FOGGIE stellar halo that has an extended in situ population is Squall, and the vast majority of these star particles were originally in the disk or inner halo and were perturbed by a single event. At $z \approx 0.7$, Squall undergoes a 2:1 prograde merger and the final coalescence of the two galaxies propels a shell of star particles that originally formed in Squall's disk into the stellar halo. The edge of this shell is clearly visible at $r \approx 230$ kpc in Figure \ref{fig:insitu}, where Squall's in situ contribution drops from a nearly constant value of 13--16\% to below 10\%. The in situ star particles perturbed during this merger, as well as a number that formed in the merging galaxy, also make significant contributions to the shell structures visible around Squall in Figure \ref{fig:haloim}. 

That being said, there is no radius at which Squall's halo is in situ-dominated. The same is true of Maelstrom, and Tempest and Hurricane are accretion-dominated beyond 6\,kpc and 11\,kpc, respectively. Even Blizzard, which has the most substantial in situ contribution, is only in situ-dominated out to 16\,kpc. Given the generally low contribution that in situ populations make to the FOGGIE stellar halos, it is somewhat unsurprising that they have little influence on the metallicity/color gradients that we derived in Section \ref{metcol}. The in situ contribution to the stellar halo along the minor axis is consistently substantially lower than the spherical average---in these simulations and in others \citep[e.g.,][]{Monachesi2016sims}---so the negative metallicity gradients that we measure for Tempest and Maelstrom trace radial variations in the accreted population, rather than a transition from a metal-rich in situ-dominated population to a metal-poor accretion-dominated one. Although the more substantial in situ populations of Blizzard and Hurricane do cause the total metallicity profile to deviate slightly from that of the accreted star particles, the metallicities of the in situ and ex situ star particles are not different enough to produce a gradient. 

The metallicities of the in situ and ex situ star particles in the inner portions of Squall's minor axis wedge are also quite similar. However, while we do not observe a color/metallicity gradient in Squall's inner halo, the left-hand panel of Figure \ref{fig:cmcomp} shows that the metallicity of Squall's stellar halo is slightly elevated relative to the relation followed by the other FOGGIE galaxies. This is due to Squall's relatively late major merger and reflects the fact that, at $r=30$\,kpc (the location at which we measure the metallicity), $>$50\% of the stars along Squall's minor axis formed in either Squall or the massive galaxy with which it merged and are thus comparatively metal-rich. Elevated stellar halo metallicity may therefore be an indication that a galaxy has experienced a major merger since $z=1$, even in cases where the halo and disk otherwise show no obvious signs of disturbance at $z=0$.

The fact that Squall's extended in situ halo population results from a late major merger suggests that the mass accretion histories of the other FOGGIE galaxies are at least partially responsible for their more centrally concentrated in situ distributions. In \S\,\ref{sec:halos}, we noted that all of the FOGGIE galaxies were selected to have completed their last major merger prior to $z\approx2$ in order to mimic the Milky Way. While Squall's final major merger was delayed in the production simulation, the other FOGGIE galaxies assembled the majority of their mass at relatively high redshifts. Their disk and inner halo regions therefore haven't been substantially perturbed since the halos themselves were much smaller and the in situ populations thus remain centrally concentrated. This is consistent with the findings of \citet{Rey2022}, who used genetically modified simulations to show that later, more violent major mergers scatter in situ stellar halo populations outward.

However, variations in mass accretion history cannot fully account for how compact the in situ stellar halo populations of the FOGGIE galaxies are compared to those in other simulations. Although large in situ populations seem to be most common in simulated galaxies with recent major mergers, many simulations of Milky Way-mass galaxies have extended in situ populations, even when the central galaxy has had a quiescent merger history \citep[e.g.,][]{Monachesi2019,Font2020,Yu2020}. This discrepancy is likely partially due to differences in feedback and star formation prescriptions \citep[e.g.,][]{Font2020}. As noted in \S\,\ref{caveats}, the FOGGIE simulations employ relatively simple routines that tend to produce small disks with tightly bound stars. However, some of the difference that we see may also be the result of our high temporal resolution. Because the snapshots that we use to identify the galaxy in which a star particle formed are only $\sim$5 Myr apart, it is less likely that we will misclassify ex situ star particles that formed during pericentric passages or in soon-to-be-stripped ram pressure compressed gas as in situ. The distributions of in situ stars in the FOGGIE halos seem to be most similar to those found by \citet{Pillepich2015} in the Eris simulation, which also has closely spaced snapshots ($\sim$30 Myr apart). In Eris, in situ star particles cease to dominate the mass budget beyond 10\,kpc and fall below 5\% of the mass at $r>30$\,kpc---similar to what we see in the halos of Tempest and Maelstrom.

More compact in situ stellar populations, like those found in FOGGIE and Eris, appear to be favored by observations. \citet{Naidu2020} find that there is a substantial in situ contribution to the Milky Way's stellar halo only within the inner 15\,kpc. Additionally, a number of authors have found that stellar halo simulations more closely match observations when in situ populations are entirely absent: \citet{Harmsen2017} found that the \citet{Bullock2005} stellar halos, which include no in situ component, are consistent with the GHOSTS stellar halos and \citet{Monachesi2019} showed that the \textsc{Auriga} stellar halos could be brought into closer agreement with GHOSTS by eliminating their in situ populations. In their exploration of the ``missing outskirts'' problem (see \S\,\ref{SBmaps}), \citet{Merritt2020} experimented with a variety of changes to the TNG100 galaxies to make them consistent with the DNGS sample and found that one of the most effective methods was reducing the spatial extent of the in situ halo. The relatively compact in situ populations of the FOGGIE stellar halos may therefore provide an explanation for why we do not see substantial excess light in the halo outskirts relative to observations.

Additionally, because the in situ populations of the FOGGIE stellar halos are so centrally concentrated, they are considerably more sensitive to the method that we use to define the stellar halo than accreted stars are. Most of the simulations mentioned earlier in this subsection use kinematic selection criteria similar to ours, but were we to use a stellar halo definition that includes all non-disk stars with $r>20$\,kpc, for instance, the in situ contribution would drop substantially for every halo except Squall (down to $<$5\% for Tempest and Maelstrom and $<25$\% for Blizzard and Hurricane). Accordingly, most of the observationally-motivated stellar halo definitions that we employ in Figure \ref{fig:masscomp} exclude the majority of stellar halo mass that is contributed by in situ star particles. This may help to explain why the FOGGIE stellar halo mass fractions appear to be more consistent with observations than many other simulations are.

\subsection{The accreted halo}
\label{acchalo}
The remainder of the star particles that populate the stellar halos of the FOGGIE galaxies are star particles that originally formed in other (mostly dwarf) galaxies. The number of galaxies that contribute star particles to the FOGGIE stellar halos ranges from 14 (Squall) to 48 (Hurricane) and roughly scales with the mass of the central host. As stated in \S\,\ref{origins}, our resolution is limited by the dark matter particle mass in the simulations: we do not resolve galaxies with $M_\mathrm{vir}<$10$^9$\,M$_\odot$ and this contributes to our choice to consolidate galaxies that merge prior to infall. The numbers quoted here should, therefore, be considered a lower limit on the number of individual contributors. 

\begin{figure}
\centering
\includegraphics[trim= 0mm 0mm 0mm 0mm, clip, width=0.47\textwidth]{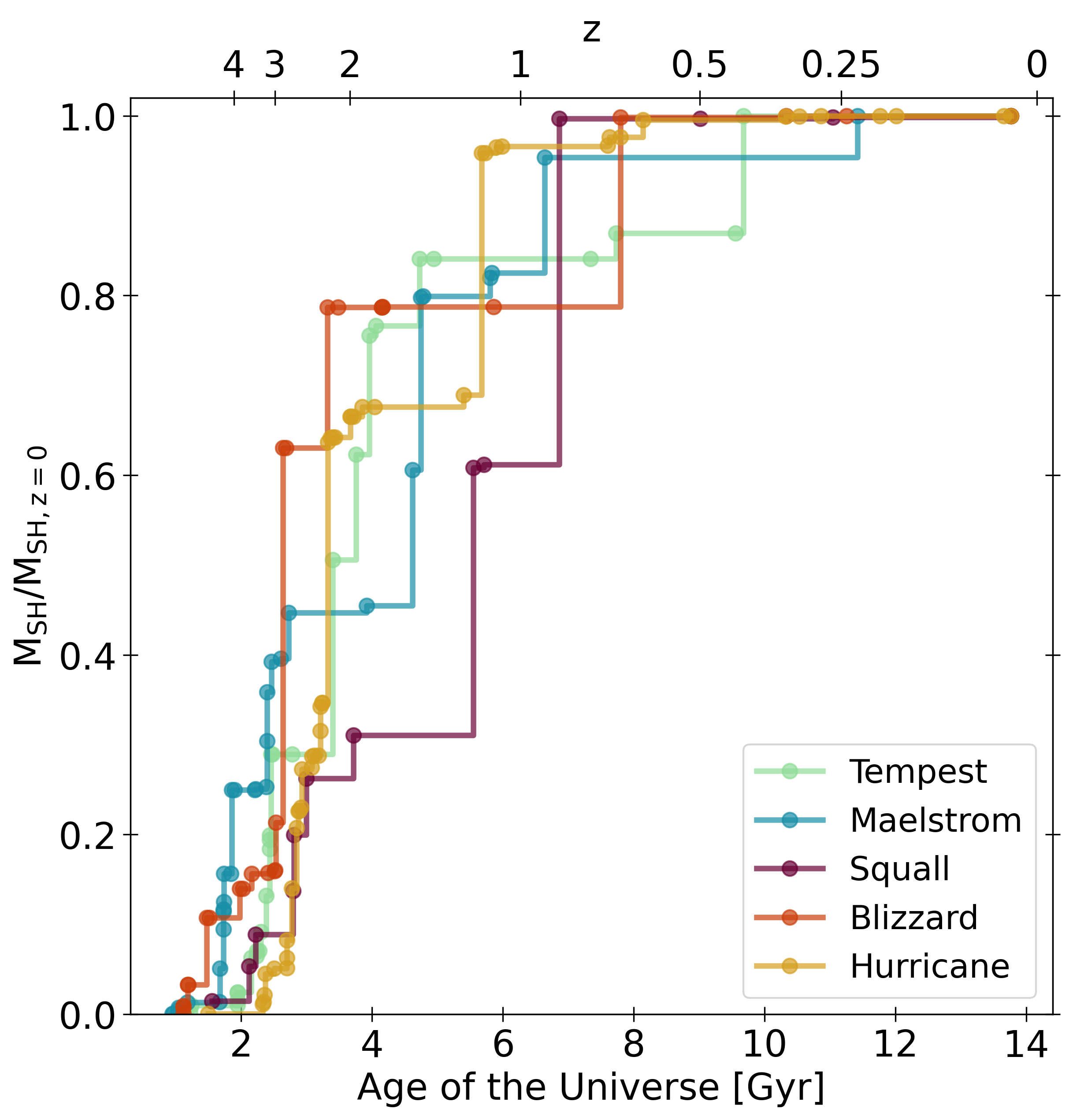}
\caption{The build-up of accreted stellar halo mass in the FOGGIE stellar halos over time. Each dot indicates an individual satellite accretion event. Only mass that exists within the stellar halo at $z=0$ is considered and the time at which the mass is accreted is assumed to be the time at which the satellite it belongs to first crosses the virial radius of the central galaxy. More massive galaxies generally build up their accreted stellar halos fastest, but all of the FOGGIE stellar halos have assembled the majority of their accreted mass by $z=1$.}
\label{fig:shgrowth}
\end{figure}

In Figure \ref{fig:shgrowth}, we show the evolution of the accreted mass in each FOGGIE stellar halo. We plot the fraction of the total accreted stellar halo mass that exists at $z=0$ as a function of time, indicating each satellite accretion event using a dot. Note that we do not consider star particles that only temporarily contribute to the mass of the stellar halo or star particles that are still part of a gravitationally self-bound satellite core at $z=0$. Additionally, we assume that whatever mass a satellite will ultimately contribute to the $z=0$ stellar halo is instantaneously added to the stellar halo at the time at which the satellite first crosses the virial radius of the central galaxy, rather than when each individual star particle is unbound from its original host. The time at which each star particle would be considered to contribute to the stellar halo by the definition that we applied in Section \ref{kinid} is therefore typically slightly later than what is shown here, particularly in cases where the satellite continues to form stars after infall. However, we use the time of infall for the sake of practicality, given the number of individual star particles that make up each halo.

The growth of stellar halo mass broadly tracks that of total mass, although the latter tends to be more gradual---particularly at later times---presumably as a result of smooth dark matter accretion. We see a general trend of the more massive FOGGIE galaxies (Blizzard and Hurricane) building up their accreted stellar halo mass earlier than the less massive ones (Tempest and Maelstrom). However, there is considerable scatter at any given time. For instance, Maelstrom is the first galaxy to amass more than 20\% of its accreted stellar halo while Hurricane is the last. All of the FOGGIE galaxies have assembled half of their final accreted stellar halo mass by $z=1$, with nearly all of the remaining assembly taking place by $z=0.6$. The relatively early growth of our stellar halos is likely largely due to our deliberate selection of galaxies that complete their last major merger at $z\gtrsim2$. Notably, Squall, the only halo which does not fulfill this criterion, acquires most of its accreted stellar halo mass later than the other galaxies. \citet{Deason2016}, who do not take merger history into account when selecting their simulated galaxies, see a much wider variety in the growth histories of their stellar halos (cf. their Figure 4).

While all of the FOGGIE galaxies assemble the first $\sim 15$\% of their accreted stellar halo mass from a relatively large number of low-mass dwarfs, there is considerable variation in where the remaining accreted mass comes from. We  see hints of a mass trend: Tempest and Maelstrom's stellar halos are primarily built from many smaller accretion events, while the mass that makes up the stellar halos of Blizzard and Squall is heavily dominated by star particles from 1 or 2 more substantial satellites. Hurricane has the highest number of contributors to its stellar halo overall, but much of its mass still comes from just a few of them. 

\begin{figure}
\centering
\includegraphics[trim= 0mm 0mm 0mm 0mm, clip, width=0.47\textwidth]{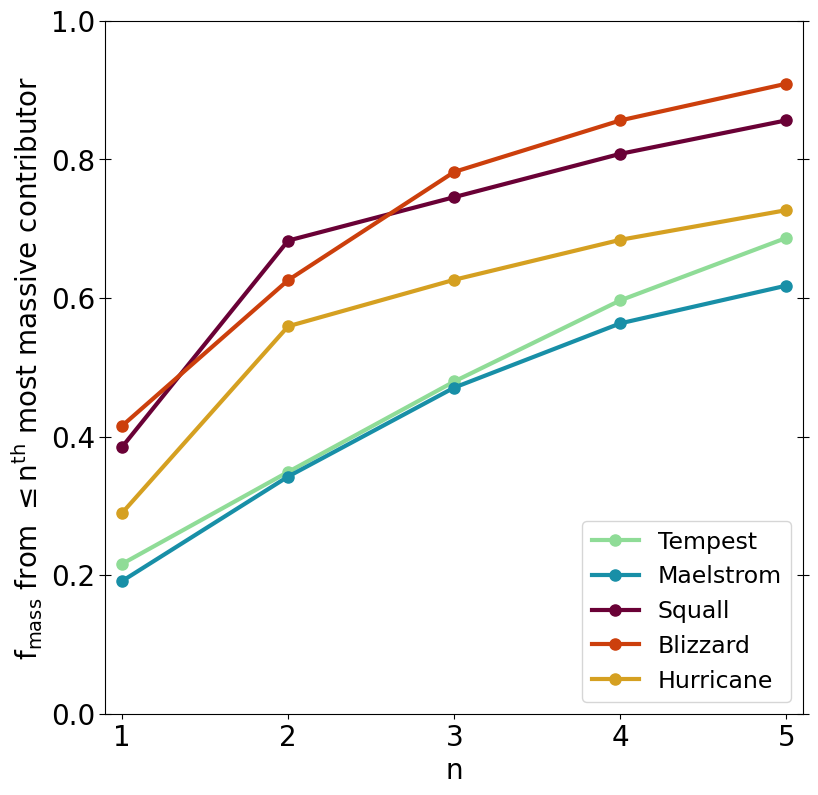}
\caption{Cumulative fraction of total accreted stellar halo mass contributed by the 1st--5th most significant contributors to each stellar halo. Although Squall and Blizzard are the most dominated by one to two accretion events, all five FOGGIE halos receive 60--90\% of their total accreted stellar halo mass from their five most significant contributors.}
\label{fig:contmass}
\end{figure}

We show the main sources of accreted stellar halo mass more clearly in Figure \ref{fig:contmass}, where we plot the cumulative fraction of accreted mass that comes from each stellar halo's five most significant contributors. While no stellar halo gets more than 50\% of its accreted mass from a single satellite, Squall, Blizzard, and Hurricane all acquire the majority of their mass in just two accretion events. If we include the five most significant contributors to each stellar halo, we can account for 60--90\% of the accreted mass of all five FOGGIE stellar halos. The predominance of stars from just a few accreted satellites is consistent with findings from previous simulations \citep[e.g.,][]{Bullock2005,Abadi2006,Deason2016} and estimates from observations \citep[e.g.,][]{Belokurov2018,Naidu2020}. \citet{Deason2016}, in particular, find that the dominant contributors to stellar halos are typically one to two relatively massive dwarfs with M$_\star =10^{8-10}$\,M$_\odot$. Although this is similar to what we see in FOGGIE, our stellar halos typically have a larger number of significant contributors than e.g., \citet{Deason2016} or \citet{D'Souza2018}. Our results are more in line with \citet{Monachesi2019}, who find that the median number of satellites required to make up 90\% of the accreted stellar halo mass in the \textsc{Auriga} simulations is 6.5---similar to the 8 we find in FOGGIE. This may reflect the fact that both FOGGIE and \textsc{Auriga} tend to overproduce stars, leading to flatter halo occupation. 

\begin{figure}
\centering
\includegraphics[trim= 0mm 0mm 0mm 0mm, clip, width=0.42\textwidth]{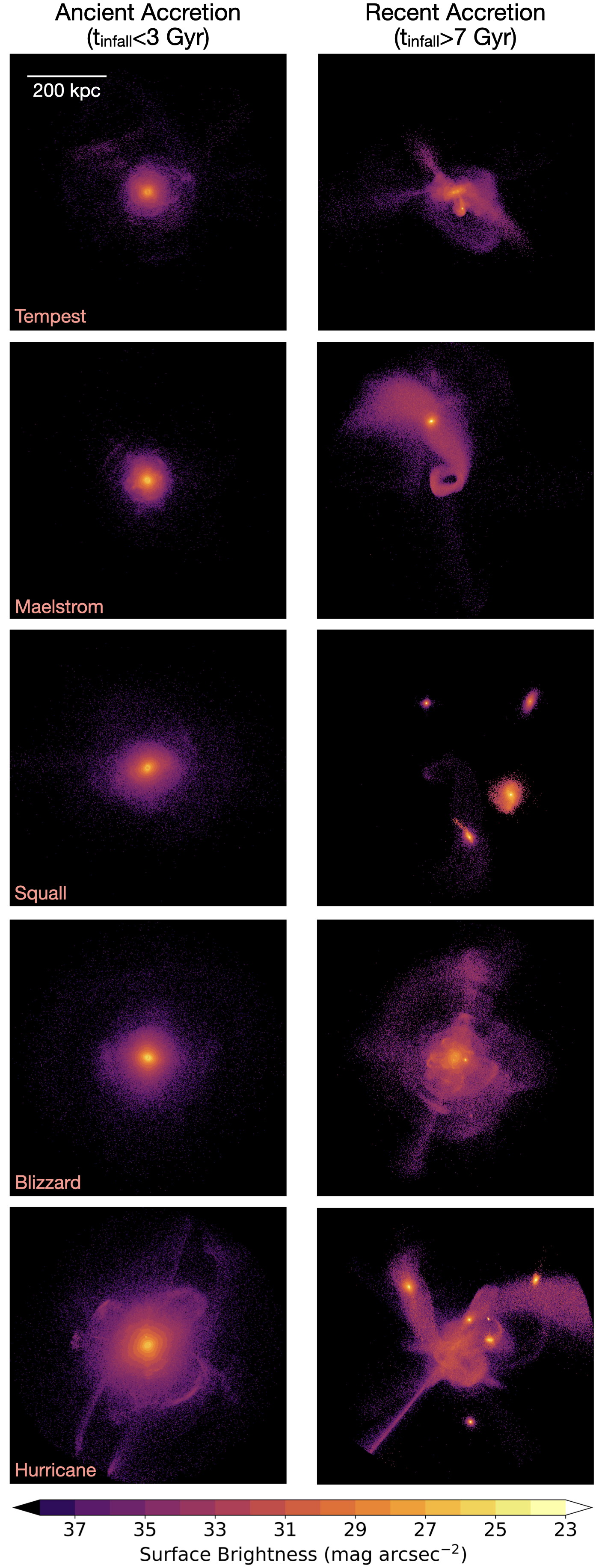}
\caption{Surface brightness maps of each FOGGIE halo divided into $g$-band light from ancient accretion events ($t<3$ Gyr; \textit{left}) and light from more recent accretion events ($t>7$ Gyr; \textit{right}), including satellite cores. Debris from early accretions tends to be centrally concentrated and well mixed while debris from more recently accreted satellites is more radially extended and is often still spatially distinct.}
\label{fig:tacc}
\end{figure}

The star particles that make up the accreted portions of stellar halos are not uniformly distributed at $z=0$. As we noted in Section \ref{metcol}, the mass and orbit of an infalling satellite has a strong influence on where its stars ultimately end up. Dynamical friction is proportional to $M_\mathrm{sat}^2$ \citep[e.g.,][]{Binney1987}, so more massive satellites will tend to sink more deeply into the gravitational potentials of their hosts and therefore deposit their stars at smaller radii \citep[e.g.,][]{Amorisco2017}. Another significant factor is the time at which a satellite is accreted. At earlier times, the central galaxy (and its dark matter halo) are smaller, leading to inside-out growth of the stellar halo \citep[e.g.,][]{Bullock2005,Font2006chem,Johnston2008,Font2011,Pillepich2014,Amorisco2017,Horta2023}. We show the influence of accretion time on the contributors to the FOGGIE stellar halos in Figure \ref{fig:tacc}, where we divide each stellar halo up into debris from its ancient ($t_\mathrm{infall}>3$ Gyr) and recent ($t_\mathrm{infall<}7$ Gyr) accretion events \citep[cf.\ Figure 16 of][]{Johnston2008}. Note that we include only star particles classified as belonging to either the ex situ stellar halo or a satellite embedded within it (i.e., with $r<350$\,kpc). Star particles that were accreted at earlier times tend to be more centrally concentrated than those that fell in later. Additionally, debris from earlier accretion tends to be more phase-mixed than debris from more recent accretion, which frequently includes still-bound satellite cores. This is due in part to the fact that the former has had more time in which to phase-mix and is located within a higher density area where dynamical times are relatively short. However, the rapid nonadiabatic growth of the central galaxy at early times also leads to faster phase-mixing \citep[e.g.,][]{Panithanpaisal2021}. Accordingly, we are most likely to find spatially distinct debris, like streams, at large radii.

Differences in concentration and orbital circularity have also been found to have more minor influences on the radial distribution of a satellite's debris \citep[e.g.,][]{Johnston2008,Amorisco2017}. More concentrated satellites will be more affected by dynamical friction, while satellites on more eccentric orbits will experience more rapid orbital decay, so stars from a more concentrated dwarf and/or a dwarf on a highly radial orbit are more likely to wind up at small radii.

\begin{figure}
\centering
\includegraphics[trim= 0mm 0mm 0mm 0mm, clip, width=0.47\textwidth]{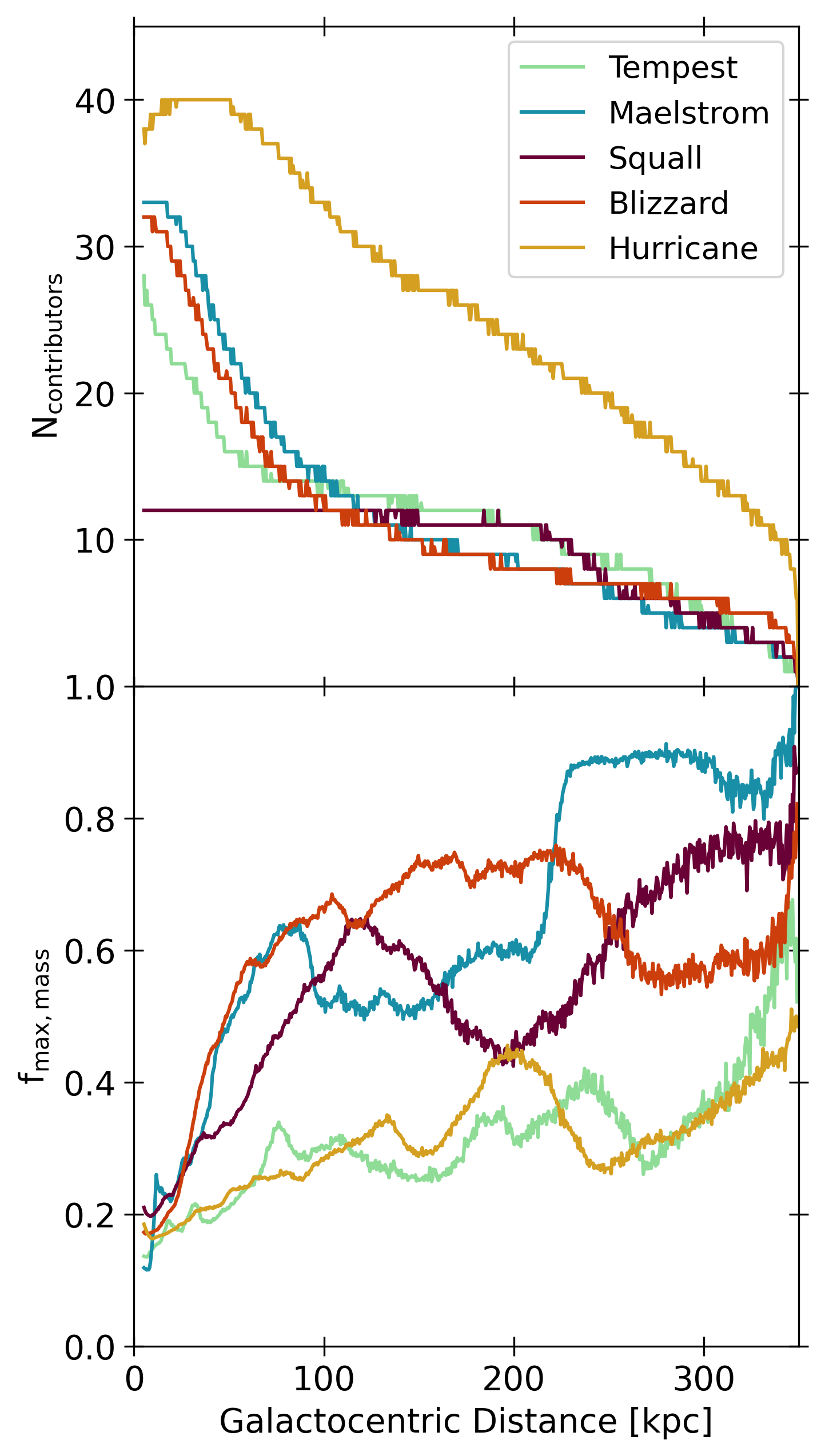}
\caption{\textit{Top}: The median number of dwarfs that contribute star particles to a given annulus within a stellar halo as a function of radius. \textit{Bottom}: The median fraction of accreted mass within an annulus that comes from the dominant dwarf contributor to that annulus. Many dwarfs contribute to the stellar halo at low galactocentric distances, but the outskirts of halos tend to be dominated by a small number of dwarfs.}
\label{fig:ncont}
\end{figure}

In Figure \ref{fig:ncont}, we show how the contributors to the FOGGIE stellar halos vary with galactocentric distance. In the top panel, we plot the median number of accreted galaxies that contribute star particles to a given annulus within the stellar halo, based on 120 random orientations and annuli spaced 0.5\,kpc apart. As we might expect, the number of contributors peaks at small radii, where the stellar halo is most concentrated and phase-mixed, and where our line-of-sight passes through a larger portion of it (although we note that the trends are the same even if we use spherical shells, rather than projected annuli). Hurricane, which has the largest total number of contributors, has the highest peak by a significant margin (40 galaxies contributing to a single annulus), while Squall, which has the smallest total number of contributors, has the lowest. The number of galaxies contributing to the halos of Tempest, Maelstrom, and Blizzard drops off rapidly after the central peak, then begins to even out at $r\approx50$\,kpc for Tempest and $r\approx80$\,kpc for Maelstrom and Blizzard. These distances roughly correspond to the outer edge of the ancient, phase-mixed debris that we see in the left-hand panel of Figure \ref{fig:tacc}. We see more gradual drop-off in Hurricane, which has a more extended phase-mixed inner halo. The number of galaxies that contribute to Squall's halo remains constant at 10--12 out to 200\,kpc. This is due to the fact that Squall does not have any contributors whose star particles have stayed exclusively at small radii. The merger that caused many of the in situ star particles that populated Squall's disk and inner halo to scatter outward also perturbed many of the ex situ star particles that contributed to the same regions; note that the location where we finally see the number of contributors fall off is also where we see the fraction of in situ star particles decrease in Figure \ref{fig:insitu}.

In the bottom panel of Figure \ref{fig:ncont}, we plot the median fraction of accreted mass within an annulus that originates from the dominant contributor to that annulus ($f_\mathrm{max,mass}$), again based on 120 random orientations of each halo. The values in this panel are generally inversely proportional to those in the top panel, since more contributors to a given annulus means that each individual galaxy contributes a smaller fraction of the mass. Accordingly, $f_\mathrm{max,mass}$ is typically low ($\sim 0.2$) at the center of the halo and increases towards 1 in the outskirts. However, this is not absolute: Blizzard and Maelstrom have nearly identical distributions of contributors, but have opposite trends in fraction of mass contributed for $r>200$\,kpc, while Tempest and Hurricane have nearly identical $f_\mathrm{max,mass}$ profiles, but vastly different numbers of contributors to their stellar halos. 

Taken together, the top and bottom panels of Figure \ref{fig:ncont} show the dual roles that accretion time and dynamical friction play in the location of debris from infalling satellites. Low mass satellites are less affected by dynamical friction and can therefore deposit their debris at large radii even at fairly early times, but they are unlikely to dominate the mass budget in the outskirts if a more massive satellite has fallen in more recently. Maelstrom is clearly illustrative of this: the mass in its outskirts is almost exclusively from the recent accretion of the LMC-mass satellite that we see debris from in the right-hand panel of Figure \ref{fig:tacc}. Blizzard has two fairly significant contributors to its outskirts, leading to lower $f_\mathrm{max,mass}$ values, while both Tempest and Hurricane have a larger number of more equal-mass contributors at large radii, resulting in relatively low $f_\mathrm{max,mass}$, even at r=300\,kpc.

\section{Implications for Future Wide-Field Surveys}
\label{futsur}
While we have been able to learn a lot about the merger histories and past dwarf companions of the Milky Way and M31 by studying their stellar halos in depth, stellar halo studies outside of the Local Group are more limited. Ground-based observations of stellar halos performed with wide-field instruments like those on Subaru \citep[e.g.,][]{Barker2009,Mouhcine2010,Barker2012,Okamoto2015,Smercina2020,Smercina2023,Gozman2023}, Magellan \citep[e.g.,][]{Bailin2011,Crnojevic2016}, and Isaac Newton \citep[e.g.,][]{Ferguson2002} have used resolved star counts of RGB stars to map the stellar densities of stellar halos and estimate their masses, luminosities, metallicities, and accretion histories. Although these observations can reach equivalent surface brightnesses as faint as $\mu_{\mathrm{V}}$=33 mag arcsec$^{-2}$ \citep{Smercina2020}, they are limited at the faint end by the increasing difficulty of star-galaxy separation \citep[e.g.,][]{Smercina2023}. Other stellar halo observations have typically been restricted to either pencil-beam surveys with telescopes like \textit{HST} \citep[e.g.,][]{Mouhcine2005,Harmsen2017}---sometimes used in conjunction with wide-field ground-based observations---or low resolution integrated light studies with instruments like the Dragonfly Telephoto Array \citep[e.g.,][]{Merritt2016,Gilhuly2022}. However, over the next decade, the astronomical community will commission a number of instruments that will combine wide fields-of-view with high resolution, making it possible for us to study large numbers of stellar halos in considerable detail for the first time. In this section, we explore how we can apply our findings to future stellar halo observations with a focus on three of these new observatories: the Vera C. Rubin Observatory, \textit{Euclid}, and the \textit{Nancy Grace Roman Space Telescope}. \\

\subsection{Integrated Light}
\label{intlight}
\begin{figure*}
\centering
\includegraphics[trim= 0mm 0mm 0mm 0mm, clip, width=0.97\textwidth]{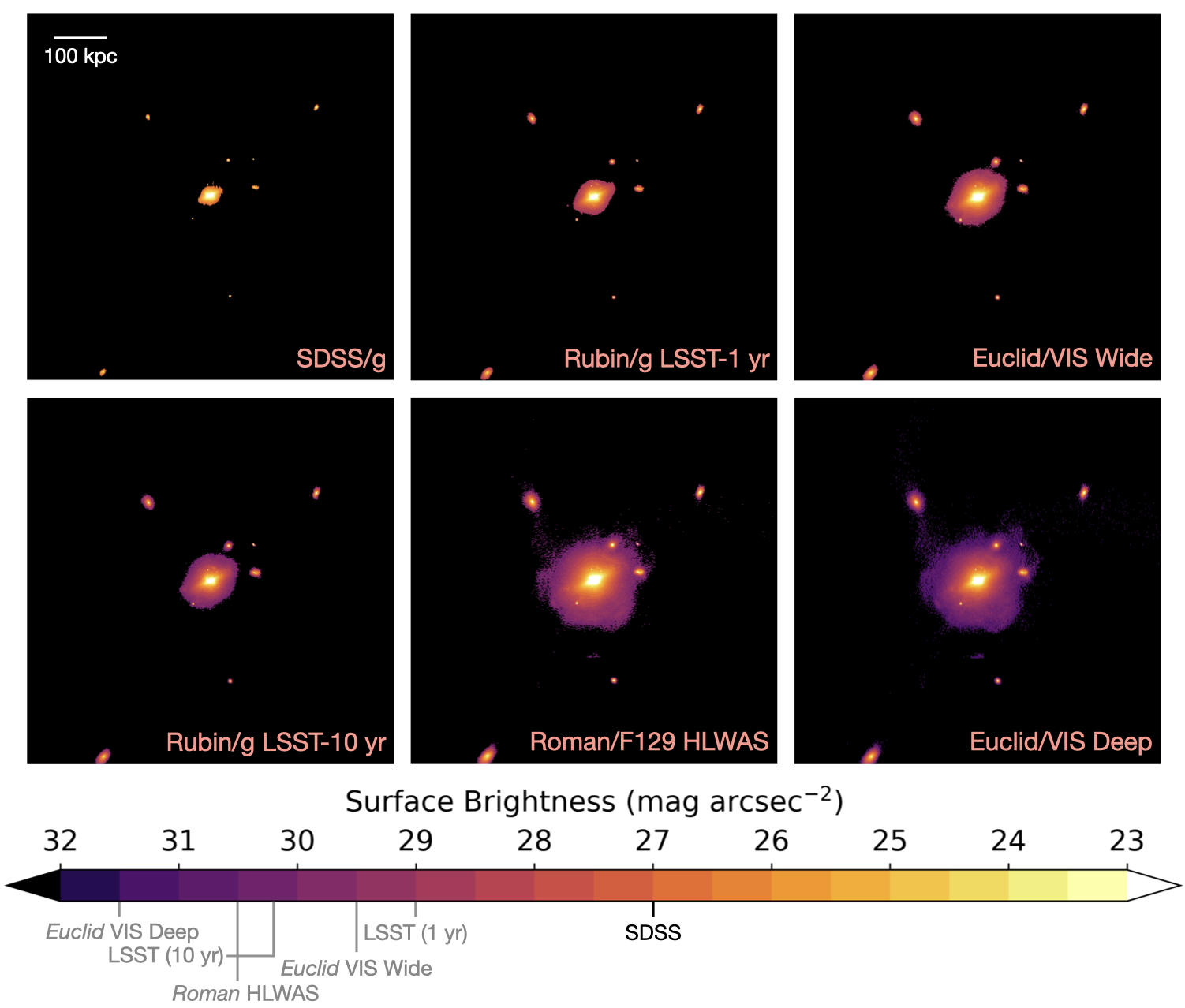}
\caption{Surface brightness maps of the FOGGIE galaxy Hurricane made using the filters and integrated light surface brightness limits of different surveys. The relevant survey and filter are listed in the bottom right corner of each image and the values and sources of the surface brightness limits are the same as those listed in Figure \ref{fig:haloim}. Any pixel with a surface brightness below the detection limit of the survey in question is shown in black. Although much of the stellar halo is still too low surface brightness to be detected, future wide-field surveys---particularly \textit{Roman}'s HLWAS and \textit{Euclid}'s Deep Survey---will allow us to probe stellar halos in far greater numbers than ever before.}
\label{fig:mocksurvey}
\end{figure*}
We will start in the limit of unresolved stars, where we observe stellar halos in integrated light. In Figure \ref{fig:mocksurvey}, we show surface brightness maps of Hurricane using the filters and integrated light surface brightness limits of different surveys. Each map is constructed in the same way as Figure \ref{fig:haloim}: Hurricane is oriented such that its central disk is edge-on and the luminosity of each star particle in a given filter is calculated with MIST models and FSPS. Any pixels with surface brightness values fainter than the 3\,$\sigma$ 10$\times$10 arcsec$^2$ surface brightness limit of the survey are shown in black to indicate that the light they contain would not be detected.

Five of the six maps we show are for future surveys. However, in the upper left, we also include a mock SDSS map for comparison, using its $g$-band filter and a surface brightness limit of 27 mag\,arcsec$^{-2}$ \citep{Pohlen2006}. The other maps use the parameters of the Rubin Observatory's Legacy Survey of Space and Time (LSST) in the $g$-band filter \citep[$\mu_\mathrm{lim} =29$ mag\,arcsec$^{-2}$ for 1-year data, and $\mu_\mathrm{lim} = 30.3$ mag\,arcsec$^{-2}$ for 10-year data;][]{Yoachim2022}, \textit{Euclid}'s Wide and Deep Surveys in the VIS filter \citep[$\mu_\mathrm{lim}=29.5$ \& 31.5 mag\,arcsec$^{-2}$, respectively;][]{EuclidCollaboration2022}, and \textit{Roman}'s High Latitude Wide Area Survey (HLWAS) in the F129 filter \citep[$\mu_\mathrm{lim}=30.5$ mag\,arcsec$^{-2}$;][]{Martinez-Delgado2023,Montes2023}. 

While the SDSS map shows only the central disk of Hurricane and the bright cores of its satellites, LSST and \textit{Euclid}'s Wide Survey reveal Hurricane's inner halo and slight asymmetries in the outskirts of its satellites that might indicate tidal disruption. \citet{Shipp2023} similarly find that many of the satellites of Milky Way-like galaxies in the FIRE simulations have tidal tails that would have gone undetected by previous surveys, but which should be detectable with Rubin and \textit{Euclid}. Both \textit{Roman}'s HLWAS and \textit{Euclid}'s Deep Survey probe Hurricane's halo out to $\sim 100$ kpc and detect multiple stellar streams. If we compare Figure \ref{fig:mocksurvey} to the deeper map of Hurricane in Figure \ref{fig:haloim}, we can see that much of the halo still lies below the detection limits of all of these surveys; observing the full complexity of the halo requires deeper observations ($\mu>33$ mag\,arcsec$^{-2}$). However, the greater depth of these surveys---particularly those that reach $\mu>$30 mag\,arcsec$^{-2}$---will yield considerable information about the infall times and orbits of the many dwarfs that contribute to stellar halos and therefore about the accretion histories of these systems. 

\subsection{Resolved Stellar Populations}
\label{resstars}
In the more local universe, we will be able to use wide-field surveys to resolve individual stars in stellar halos in far greater numbers than ever before. Previous resolved star surveys of stellar halos beyond the Local Group have typically been limited to either small fields because a telescope like \textit{HST} was required to detect and resolve the faint halo population or to the brightest giant stars because only ground-based telescopes had wide enough fields-of-view to capture entire stellar halos. However, high-resolution wide-field telescopes will enable us to acquire detailed panoramic data more akin to what surveys like PAndAS have achieved with M31's halo. 

Because they are space-based, both \textit{Roman} and \textit{Euclid} are expected to have better galaxy-star separation than Rubin, which is limited by atmospheric seeing \citep[although see, e.g.,][]{Mutlu-Pakdil2021,Martin2022}. However, while \textit{Euclid}'s resolution in its optical bands is very similar to that of \textit{Roman}, its NIR resolution is slightly worse. This is particularly relevant for older stellar populations, like those that make up the vast majority of the stellar halo, because their SEDs peak in the NIR \citep[e.g.,][]{Martinez-Delgado2023}, making them easier to observe. We can see this effect in Figure \ref{fig:mocksurvey} when comparing the surface brightness map in \textit{Euclid}'s VIS Deep Survey to that in \textit{Roman}'s F129 filter. Although the VIS Deep Survey is expected to probe 1 mag\,arcsec$^{-2}$ deeper than \textit{Roman}'s HLWAS, the two maps are almost identical. This occurs, not because there are few features with surface brightnesses in between 30.5 and 31.5 mag\,arcsec$^{-2}$, but because the stellar populations that make up the stellar halo are brighter in the NIR. A \textit{Roman} F129 surface brightness map with $\mu_\mathrm{lim}=31.5$\,mag\,arcsec$^{-2}$ shows at least two streams that are not visible in the \textit{Euclid} VIS Deep map. We will therefore primarily focus on \textit{Roman} for this part of the discussion, although it is worth noting that, if the footprint of the \textit{Roman} HLWAS overlaps with LSST or \textit{Euclid}'s surveys, we will likely be able to glean more information about observed stellar populations than we could get from any individual survey \citep[e.g.,][]{Eifler2021,Gezari2022}.

We will also primarily be focusing on information that can be derived from the positions of stars and their luminosities in various filters. Although \textit{Roman} is expected to yield some kinematic data for stars in the nearby universe \citep{WFIRSTAstrometryWorkingGroup2019}, we expect to get far more precise data with the upcoming thirty meter-class telescopes \citep[e.g., the Giant Magellan Telescope;][]{Johns2012}. We therefore leave kinematic-based diagnostics (including orbit characterization) for a future paper. However, stellar positions and luminosities can provide a significant amount of information on their own through color-magnitude diagrams (CMDs). With deep enough CMDs, we may be able to derive ages, metallicities, and star formation histories for stellar halos. While this data provides useful information even if we can only come up with global values for the stellar halo (see, e.g., Figure \ref{fig:cmcomp}), we may also be able to use it to identify debris from different contributors and infer the properties of their progenitors.

\begin{figure*}
\centering
\includegraphics[trim= 0mm 0mm 0mm 0mm, clip, width=0.97\textwidth]{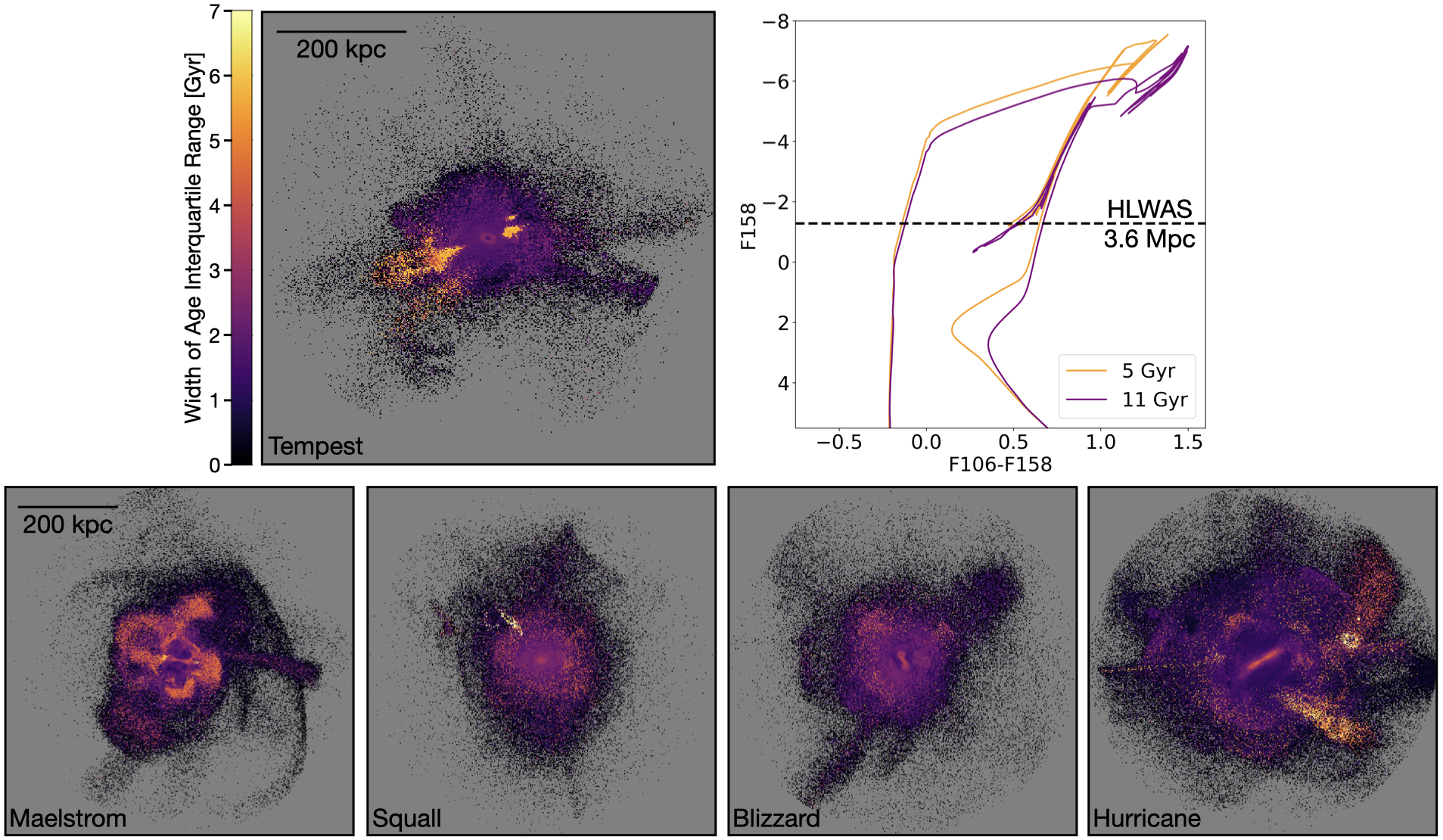}
\caption{\textit{Top}: In the left panel, we show Tempest's stellar halo, colored by the width of the age interquartile range of the star particles that contribute to each (1.5\,kpc)$^2$ pixel. Gray pixels contain no star particles. While the majority of the halo is uniformly old, there is a clear structure populated by younger stars. In the right panel, we show theoretical isochrones for this younger population (orange) and the older population (purple) that makes up the bulk of the halo. The giant branches and main sequence turn-offs are distinguishable. \textit{Bottom}: Age interquartile ranges for the other four halos. While Squall and Blizzard are almost uniformly old, Maelstrom and Hurricane both have structures with noticeable variations in age.}
\label{fig:ageIQR}
\end{figure*}

In Figure \ref{fig:ageIQR}, we explore how the ages of stars vary within individual stellar halos. In the top left panel, we show Tempest's stellar halo, with each (1.5\,kpc)$^2$ pixel colored by the width of the age interquartile range of the stellar halo star particles that fall within it. As we noted in Figure \ref{fig:kinid}, the bulk of Tempest's stellar halo is very old (age$>$10\,Gyr) with variations on the order of $\pm$1\,Gyr. As observational uncertainties concerning the age of a star correlate with the value of that age \citep[e.g.,][]{Weisz2011}, we cannot expect to be able to reliably distinguish a 10\,Gyr old star from an 11\,Gyr old one, so this offers little extra information. However, there are two wedge-shaped regions close to the center of the stellar halo in which the age interquartile range is $\sim$6\,Gyr. To determine whether this difference in age is detectable, we show representative theoretical isochrones from MIST in \textit{Roman} filters in the adjacent panel. In orange, we show the isochrone for a stellar population with an age of 5\,Gyr---the median age of the star particles that make up these structures---and in purple we show  the isochrone for a stellar population with an age of 11\,Gyr---the median age of the underlying stellar halo. The separation between the two isochrones in the regions surrounding the giant branch and the main sequence turn-off are considerable and it is therefore likely that the younger stars populating these regions could be identified as belonging to a distinct structure---in this case debris from a late infalling satellite.

The expected 5$\sigma$ depth of \textit{Roman}'s HLWAS is $\sim\,26.5$ \citep{Montes2023}---shown as a dashed line on the CMD at the distance of M81---so we will not detect the main sequence turn-off for distances greater than $\approx$1 Mpc. We note, though, that deeper measurements are possible with General Observer (GO) programs. Additionally, the HLWAS is expected to resolve individual RGB stars out to $\sim$10 Mpc \citep{Lancaster2022}, with the tip of the red giant branch and bright asymptotic giant branch (AGB) stars distinguishable to even greater distances. \citet{Harmsen2023} show that the ratio of AGB to RGB stars alone can be used to constrain the age of a population in a stellar halo. It is therefore possible that structures in age-space will prove to be a key component of \textit{Roman}'s survey of stellar halos throughout the Local Volume.

In the bottom panels of Figure \ref{fig:ageIQR}, we show the age interquartile ranges for the other four halos. Although Blizzard and Squall have nearly uniformly old stellar halos, Maelstrom and Hurricane both have structures with diverse enough ages that they are likely to be detectable within a CMD. Like Tempest, Maelstrom has a single dwarf contributor that produces most of this structure, but Hurricane's age diversity comes from at least 4 different recently accreted dwarfs. By combining these anomalies in age-space with position data, we can pick out structures that likely originated from the same dwarf contributor and potentially reconstruct the infall time, orbit, and star formation history of that dwarf. It is worth noting that observations will likely also reveal substructures in metallicity similar to those already discovered in the Milky Way and M31 \citep[e.g.,][]{Cohen2018,Naidu2020}. However, as the FOGGIE simulations do not reliably model metallicity, we will not attempt to make any predictions about this possibility here.

In regions where there are no discernible structures in age- or metallicity-space, we may be compelled to rely largely on position data. Fortunately, certain types of debris, like streams, can remain intact and spatially distinct for long periods of time \citep[e.g.,][]{Johnston1996,Pearson2015}. While many of the stream-finding algorithms that have been developed for use with \textit{Gaia} data are designed to take advantage of the 6D phase-space information available in much of the Milky Way's stellar halo \citep[e.g.,][]{Malhan2018,Shih2022}, other algorithms, such as the \textsc{Hough Stream Spotter} \citep{Pearson2019,Pearson2022}, work by looking for linear structures in stellar position data. Such streamfinders can (and have) been used to identify structures in stellar halos around external galaxies and will likely prove to be crucial in the quest to disentangle structures in distant stellar halos.

\begin{figure*}
\centering
\includegraphics[trim= 0mm 0mm 0mm 0mm, clip, width=0.97\textwidth]{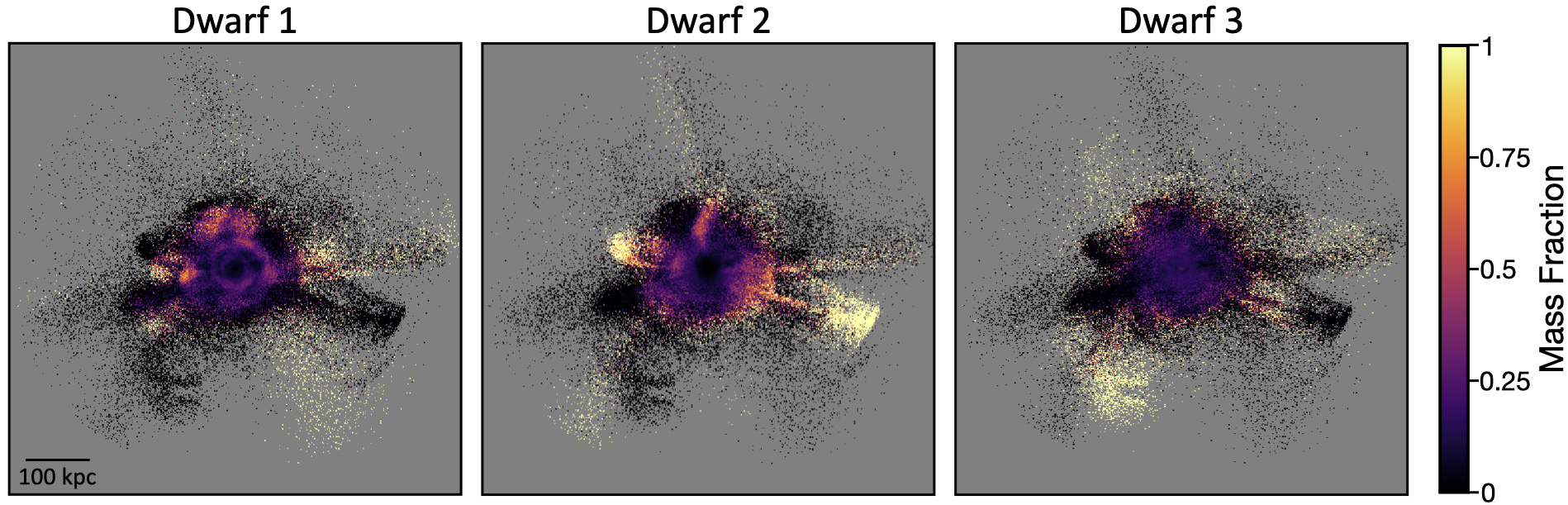}
\caption{Fraction of mass contributed to each (1.5\,kpc)$^2$ pixel by three different dwarfs along the same line of sight in Tempest's stellar halo. Gray pixels contain no star particles. Although most of the debris does not form linear structures from this line-of-sight and is therefore unlikely to be identified with a stream-finding algorithm, it still remains clumped together, particularly in the outskirts of the stellar halo.}
\label{fig:dwarfs}
\end{figure*}

Even when debris does not form structures that look like streams along a given line-of-sight, stars from individual satellites---particularly from either low mass or recently accreted objects---are often still spatially distinct. In Figure \ref{fig:dwarfs}, we show examples of debris from three different dwarf galaxies that contribute to Tempest's stellar halo. Each panel shows the same line-of-sight on the entire $z=0$ halo and each (1.5\,kpc)$^2$ pixel is colored by the fraction of halo mass within it that comes from star particles that formed in each individual dwarf. The dwarfs are arranged in order of latest to earliest infall time. Even though Dwarf 2 was accreted $\approx$9\,Gyr ago, much of its debris is still thin and structured enough to be identified either by eye or with a traditional stream-finding algorithm. By contrast, the debris from Dwarfs 1 and 3 appears to be relatively diffuse and would likely be difficult to disentangle with typical stream-finders. Dwarf 3 is a fairly massive dwarf that was accreted $>$10\,Gyr ago and is therefore largely phase-mixed, but Dwarf 1 fell in quite recently ($\approx$4\,Gyr ago) and appears more structured when viewed from other lines-of-sight. However, much of the debris from theses dwarfs is still spatially clustered, to the extent that there are regions, particularly in the outskirts of Tempest, where the stellar halo is dominated by debris from just one of them. This is consistent with results from \citet{Font2008}, who find that bright features in the outer halos of the \citet{Bullock2005} simulations tend to originate from a single dwarf.

We have already demonstrated the clustering of debris in halo outskirts, in a more general sense, in the bottom panel of Figure \ref{fig:ncont}: three of the five FOGGIE stellar halos get more than 50\% of their mass within a given annulus from a single contributor at most galactrocentric distances beyond 100 kpc. However, looking at Figure \ref{fig:dwarfs}, we can see that Figure \ref{fig:ncont} fails to take into account how azimuthally clustered much of the debris from a single dwarf may be. The logical next step is to develop a more observationally-motivated version of Figure \ref{fig:ncont}, which uses the 2D positions of star particles (as they appear when projected onto an arbitrary plane) to determine how large of an area is typically dominated by debris from a single dwarf -- an ``optimal search area". We could then potentially use this information to place statistical constraints on the properties of the dwarfs that contributed to an observed stellar halo, even in regions where we cannot pick out individual structures in either age- or position-space. This sort of measurement would, however, be impacted by the use of discrete star particles in our simulations. We assign a single position and velocity to each star particle, but, in truth, each star that a particle represents should be spread over a phase-space volume, the size of which is likely impacted by a number of factors. In the next phase of this research, we plan to generate more realistic synthetic data by using software like \textsc{ananke} \citep{Sanderson2020} and \textsc{GalaxyFlow} \citep{Lim2022} to better model the small-scale structure of our stellar halos so that we can more effectively test the efficacy of tools like streamfinders on simulated data and use the FOGGIE simulations to identify optimal search areas for telescopes like \textit{Roman}.

\section{Summary}
\label{summ}
We use the FOGGIE suite, a set of high-resolution cosmological simulations of Milky Way-like galaxies run with fine time resolution and low mass star particles, to study the properties of stellar halos and the galaxies that they are built from. We summarize our primary findings below:
\begin{itemize}
\item The masses, surface brightness profiles, and metallicity/color gradients of the FOGGIE stellar halos are generally consistent with those of observed stellar halos. We see only slight evidence that the FOGGIE stellar halos may have excess light in their outskirts, like that which has been found in a number of other simulations \citep[e.g.,][]{Merritt2020,Keller2022}. We largely attribute this to the relatively compact in situ populations of the FOGGIE stellar halos.
\item Although the FOGGIE simulations were selected to cover a small range of virial mass and to have somewhat similar merger histories, their stellar halos have diverse properties. They vary considerably in appearance---from the stream- and shell-dominated halos of Hurricane and Maelstrom to the relatively smooth halos of Blizzard and Squall---and range over $\approx$3 magnitudes in surface brightness at any given radius. Many of their features are extremely low surface brightness ($\mu_g>$33.5 mag arcsec$^{-2}$) and are therefore only detectable due to FOGGIE's use of low mass ($\sim$1000\,M$_\odot$) star particles.
\item While the majority of the stars that make up the FOGGIE stellar halos originate in dwarf galaxies disrupted by the more massive central, 30-40\% are formed either in the central disk or in the halo itself. Although the overall mass fraction of in situ stars is consistent with other simulations, in situ populations in the FOGGIE simulations tend to be more centrally concentrated, in line with recent observations of the Milky Way \citep[e.g.,][]{Naidu2020}. This difference appears to be due to a combination of the high temporal resolution and conservative star formation and feedback prescriptions employed in the FOGGIE simulations and the quiescent merger histories that characterize most of the galaxies.
\item The only FOGGIE galaxy that experiences a late major merger (Squall) also has a more extended in situ population and sits higher on the M$_\mathrm{SH}$-metallicity relation than the other galaxies. High stellar halo metallicity may therefore be an indication that a galaxy has experienced a major merger at $z<1$, even when the disk and stellar halo appear undisturbed at $z=0$.
\item Each FOGGIE stellar halo contains stars that originally formed in 14 (Squall) to 48 (Hurricane) other galaxies and the majority of the mass contributed by these galaxies is accreted prior to $z=1$. The more massive FOGGIE galaxies tend to build up their accreted mass more quickly than the lower mass FOGGIE galaxies, and the bulk of this mass comes from fewer, more massive galaxies. However, the five most massive accreted objects contribute the majority (60--90\%) of the accreted mass in all five of the FOGGIE stellar halos.
\item The number of contributors to a stellar halo, the masses of those contributors, and the times at which they are accreted all play a significant role in the composition of a stellar halo at any given radius. The FOGGIE stellar halos tend to have the most contributors at small galactocentric distances and do not receive more than $\approx$20\% of their accreted mass from any individual contributor within this region. Beyond the phase-mixed inner halo, however, the number of contributors drops off and the accreted mass fraction is dominated by a single contributor at galactocentric distances $>$100 kpc in three of the five halos. The remaining two halos have a large number of more equal-mass contributors at large galactocentric distances.
\item Future surveys by high-resolution wide-field telescopes like Rubin, \textit{Roman}, and \textit{Euclid} will probe the outskirts of large numbers of galaxies to much greater depths than previous large surveys, allowing us to study the stellar halos of far more galaxies than ever before. Mock observations of the FOGGIE galaxies based on \textit{Roman}'s HLWAS and \textit{Euclid}'s Deep Survey suggest that these surveys will detect stellar halos out to $\approx$100 kpc and identify stellar streams at even larger galactocentric distances in integrated light.
\item Three of the five FOGGIE stellar halos contain structures with sufficiently diverse ages that stars belonging to them should be identifiable in CMDs made with \textit{Roman} throughout the Local Volume.  
\end{itemize}

The methods that we have explored in this paper for disassembling stellar halos in order to better understand their contributors and overall histories are truly just the tip of the iceberg. High-resolution wide-field observatories like Rubin, \textit{Euclid}, and \textit{Roman} will open up an entirely new parameter space for understanding stellar halos and the dwarf galaxies that create them. Resolving stellar populations in external galaxies is a particularly promising avenue and we are working on creating more realistic synthetic data from the FOGGIE stellar halos in order to prepare for the multitude of data that will soon be available to the astronomical community.

\section*{Acknowledgments}
The authors thank the anonymous referee for a thorough and thoughtful review that improved the content and organization of this paper. We also thank Sebastian Gomez, Ayan Acharyya, Karoline Gilbert, Eric Bell, Sarah Loebman, Claire Kopenhafer, Erik Tollerud, Marla Geha, Jillian Bellovary, and Ferah Munshi for encouragement and useful discussions related to this work. During the course of this work ACW and JT were supported by the \textit{Nancy Grace Roman Space Telescope} Project, under the Milky Way Science Investigation Team. RA, CL, and MSP were supported for this work in part by NASA via an Astrophysics Theory Program grant 80NSSC18K1105. RA and CL also acknowledge financial support from the STScI Director’s Discretionary Research Fund (DDRF). BWO acknowledges support from NSF grants \#1908109 and \#2106575 and NASA ATP grants NNX15AP39G and 80NSSC18K1105. RA's efforts for this work were additionally supported by HST GO \#16730. NB acknowledges support from the 2022 STScI Space Astronomy Summer Program. BDS is supported by Science and Technology Facilities Council Consolidated Grant RA5496.

Computations described in this work were performed using the publicly-available \textsc{Enzo} code (\href{http://enzo-project.org}{http://enzo-project.org}), which is the product of a collaborative effort of many independent scientists from numerous institutions around the world. Their commitment to open science has helped make this work possible. The python packages {\sc matplotlib} \citep{hunter2007}, {\sc numpy} \citep{walt2011numpy}, {\sc tangos} \citep{pontzen2018}, \textsc{scipy} \citep{scipy2020}, {\sc yt} \citep{ytpaper}, {\sc datashader} \citep{datashader}, and {\sc Astropy} \citep{astropy2013,astropy2018,astropy2022} were all used in parts of this analysis. Resources supporting this work were provided by the NASA High-End Computing (HEC) Program through the NASA Advanced Supercomputing (NAS) Division at Ames Research Center and were sponsored by NASA's Science Mission Directorate; we are grateful for the superb user-support provided by NAS. 

 \section*{Data Availability}
 These results were generated from the FOGGIE cosmological simulation suite. \textsc{Tangos} databases containing the global properties of each galaxy in each snapshot are available upon email request.
\bibliographystyle{aasjournal}
\bibliography{SH}
\label{lastpage}
\end{document}